\documentclass[twocolumn]{emulateapj}

\usepackage{amsmath}
\usepackage{physics}
\usepackage{graphicx}
\usepackage{color}
\usepackage[colorlinks]{hyperref}
\hypersetup{
    colorlinks,	
    citecolor=blue,
}

\newcommand{\be}{\begin{eqnarray}}
\newcommand{\ee}{\end{eqnarray}}

\shorttitle{Towards more accurate synthetic reflection spectra}
\shortauthors{Mirzaev et al.}

\begin{document}

\title{Towards more accurate synthetic reflection spectra:\\improving the calculations of returning radiation}

\author{Temurbek~Mirzaev\altaffilmark{1}, Shafqat~Riaz\altaffilmark{2,1}, Askar~B.~Abdikamalov\altaffilmark{3,1,4}, Cosimo~Bambi\altaffilmark{1,3, \dag}, Thomas~Dauser\altaffilmark{5}, Javier~A.~Garcia\altaffilmark{6,7}, Jiachen~Jiang\altaffilmark{8}, Honghui~Liu\altaffilmark{1}, and {Swarnim~Shashank}\altaffilmark{1}}

\altaffiltext{1}{Center for Field Theory and Particle Physics and Department of Physics, 
Fudan University, 200438 Shanghai, China. \email[\dag E-mail: ]{bambi@fudan.edu.cn}} 
\altaffiltext{2}{Theoretical Astrophysics, Eberhard-Karls Universit\"at T\"ubingen, D-72076 T\"ubingen, Germany}
\altaffiltext{3}{School of Natural Sciences and Humanities, New Uzbekistan University, Tashkent 100007, Uzbekistan}
\altaffiltext{4}{Ulugh Beg Astronomical Institute, Tashkent 100052, Uzbekistan}
\altaffiltext{5}{Dr. Karl Remeis-Observatory and Erlangen Centre for Astroparticle Physics, Friedrich-Alexander Universit\"at Erlangen-N\"urnberg, D-96049 Bamberg, Germany}
\altaffiltext{6}{X-ray Astrophysics Laboratory, NASA Goddard Space Flight Center, Greenbelt, MD 20771, USA}
\altaffiltext{7}{Cahill Center for Astronomy and Astrophysics, California Institute of Technology, Pasadena, CA 91125, USA}
\altaffiltext{8}{Institute of Astronomy, University of Cambridge, Madingley Road, Cambridge CB3 0HA, UK}

\begin{abstract}
We present a new model to calculate reflection spectra of thin accretion disks in Kerr spacetimes. Our model includes the effect of returning radiation, which is the radiation that is emitted by the disk and returns to the disk because of the strong light bending near a black hole. The major improvement with respect to the existing models is that it calculates the reflection spectrum at every point on the disk by using the actual spectrum of the incident radiation. Assuming a lamppost coronal geometry, we simulate simultaneous observations of \textsl{NICER} and \textsl{NuSTAR} of bright Galactic black holes and we fit the simulated data with the latest version of {\tt relxill} (modified to read the table of {\tt reflionx}, which is the non-relativistic reflection model used in our calculations). We find that {\tt relxill} with returning radiation cannot fit well the simulated data when the black hole spin parameter is very high and the coronal height and disk's ionization parameter are low, and some parameters can be significantly overestimated or underestimated. We can find better fits and recover the correct input parameters as the value of the black hole spin parameter decreases and the value of the coronal height increases.
\end{abstract}


\section{Introduction}

Blurred reflection features are common in the X-ray spectra of stellar-mass black holes in X-ray binaries and supermassive black holes in active galactic nuclei~\citep{1989MNRAS.238..729F,1995Natur.375..659T,2007MNRAS.382..194N,2009ApJ...697..900M,2013MNRAS.428.2901W}. They are generated in the strong gravity regions of black holes by illumination of a cold accretion disk by a hot corona~\citep{1989MNRAS.238..729F,2003PhR...377..389R,2021SSRv..217...65B}. X-ray reflection spectroscopy refers to the analysis of these relativistically blurred reflection features and can be a powerful technique to study the accretion process around black holes~\citep{2019SCPMA..6229504D}, measure black hole spins~\citep{2021ARA&A..59..117R,2021SSRv..217...65B}, and test Einstein's theory of general relativity in the strong field regime~\citep{2019ApJ...875...56T,2021ApJ...913...79T}.

The past decade has seen tremendous progress in X-ray reflection spectroscopy, thanks both to a new generation of reflection models and new observational facilities. Today the state-of-the-art in reflection modeling is represented by the relativistic models {\tt relxill}~\citep{2013MNRAS.430.1694D,2014ApJ...782...76G}, {\tt reltrans}~\citep{2019MNRAS.488..324I}, {\tt reflkerr}~\citep{2008MNRAS.386..759N,2019MNRAS.485.2942N}, and {\tt kyn}~\citep{2004ApJS..153..205D}, and by the non-relativistic models {\tt xillver}~\citep{2013ApJ...768..146G} and {\tt reflionx}~\citep{2005MNRAS.358..211R}. Despite the remarkable improvements with respect to the previous generation of reflection models, even these models present a number of simplifications, so caution is necessary when we attempt to infer very precise measurements of accreting black holes.

Some simplifications are relatively well studied and they do not affect significantly the final measurements of a system if we select properly the spectra to analyze. For example, current models normally assume Keplerian and infinitesimally thin accretion disks. At least in the case of fast-rotating sources, such a simple model works quite well, without introducing appreciable systematic uncertainties in the final measurements, as it was shown by theoretical studies with GRMHD simulations~\citep{2022ApJ...938...53S} as well by the analysis of specific sources with models admitting deviations from Keplerian motion and disks with finite thickness~\citep{2020ApJ...899...80A,2020PhRvD.102j3009T,2021ApJ...913..129T,2022MNRAS.514.3246J}. However, such a simple model breaks down and we can have unacceptably large systematic errors if we do not select properly the sources; for instance, if we use our reflection models to analyze sources with a mass accretion rate close to their Eddington limit~\citep{2020MNRAS.491..417R,2020ApJ...895...61R}.

The choice of the model for the emissivity profile is more subtle. If we knew the geometry and the properties of the corona, the emissivity profile could be calculated with ray-tracing techniques. For coronae of unknown geometry, it is common to employ a broken power law profile, or even a twice broken power law profile, where the values of the emissivity indices and the breaking radii can be determined by the fit. A twice broken power law profile should be able to approximate well most coronal geometries~\citep{2012MNRAS.424.1284W,2017MNRAS.472.1932G}, and in the analysis of high-quality data we should be able to infer the general behavior of the emissivity profile~\citep[see, e.g.,][]{2011MNRAS.414.1269W}.

The returning radiation (or self-irradiation) is the radiation that is emitted by the accretion disk and returns to the accretion disk because of the strong light bending near the black hole. \citet{1976ApJ...208..534C} was the first to present the calculations of the returning radiation of thermal spectra of thin disks. It turns out that a thermal spectrum in which the returning radiation is taken into account appears like a thermal spectrum without returning radiation but a slightly higher mass accretion rate~\citep{2005ApJS..157..335L}, and therefore the returning radiation is normally ignored in the spectral analysis of thermal components. On the other hand, the effect of the returning radiation is definitively important and non-negligible in the polarization of thermal spectra~\citep{2009ApJ...701.1175S}.

In the context of disk reflection, the returning radiation was first studied in \citet{1997MNRAS.288L..11D}, where the authors found that the effect is not important if the corona corotates with the disk. \citet{2008MNRAS.386..759N} confirmed this conclusion but also pointed out that the returning radiation is instead important if the corona is static and close to the black hole event horizon. The importance of the returning radiation in the case of lamppost coronae was illustrated in \citet{2016ApJ...821L...1N} and \citet{2018MNRAS.477.4269N}, where the authors showed in which conditions the effect of the returning radiation can be significant.

The impact of the returning radiation on reflection spectra has been recently investigated in a few studies with different approximations~\citep{2020MNRAS.498.3302W,2022MNRAS.514.3965D,2021ApJ...910...49R,2023arXiv230312581R}. All these studies agree that the effect of the returning radiation is important only for fast-rotating black holes ($a_* > 0.9$, assuming that the inner edge of the accretion disk is at the innermost stable circular orbit, or ISCO) and when the corona is compact and close to the black hole (i.e., when the corona illuminates mainly the very inner part of the accretion disk; for a lamppost corona $h < 5~r_{\rm g}$, where $h$ is the coronal height and $r_{\rm g}$ is the gravitational radius). These studies also agree on the fact that most measurements are not significantly affected if the theoretical model used to analyze the data does not include the effects of the returning radiation. In particular, the returning radiation does not seem to have a significant impact on the estimate of key-parameters like the black hole spin. However, all these studies employ some important simplifications in the calculation of the returning radiation and, at the same time, fast-rotating black holes with compact coronae close to them are the sources for which we can naturally get more precise measurements, as their relativistic reflection features can be stronger. Further studies are thus necessary to figure out the actual impact of the returning radiation on X-ray reflection spectroscopy measurements.

In this manuscript, we present a new model to calculate reflection spectra of thin accretion disks in Kerr spacetimes produced by lamppost coronae. Our model does not exploit the existence of the Carter constant and therefore it can calculate reflection spectra of thin accretion disks in other stationary, axisymmetric, and asymptotically-flat spacetimes with minor modifications~\citep{GRtests}. If compared to the existing models, it presents two important improvements: $i)$ the non-relativistic reflection spectrum at every point on the disk is calculated by taking into account the actual spectrum of the incident radiation, which consists of a power law spectrum from the corona and a reflection spectrum from the returning radiation, and $ii)$ the effect of the returning radiation is calculated up to the fourth order, when we find that further iterations are negligible. We note that point $ii)$ is important to infer the actual spectrum of the incident radiation at point $i)$, while it does not appreciably change the estimate of the emissivity profile.

We use the reflection spectra calculated by our new model to simulate simultaneous observations of bright Galactic black holes with \textsl{NICER} and \textsl{NuSTAR}. The simulated data are fit with the latest version of {\tt relxill}, where the effect of the returning radiation is included only in the emissivity profile generated by lamppost coronae~\citep{2022MNRAS.514.3965D}. From the spectral analysis of the simulated data, we find that we cannot obtain good fits when the value of the black hole spin parameter is very high and the values of the coronal height and of the disk's ionization parameters are low, and in such a case the estimate of some parameters can be significantly biased. The quality of the fits improves and the systematic uncertainties decrease as we decrease the value of the black hole spin parameter and we increase the value of the coronal height. If we increase the value of the disk's ionization parameter, the quality of the fits improves, but mainly because the reflection features become weak, while we can have still problems to recover the correct input parameters. Our results do not contradict previous studies on the impact of the returning radiation~\citep{1997MNRAS.288L..11D,2008MNRAS.386..759N,2016ApJ...821L...1N,2018MNRAS.477.4269N,2020MNRAS.498.3302W,2022MNRAS.514.3965D,2021ApJ...910...49R,2023arXiv230312581R}, but provide a more clear and solid picture of this issue on an important region of the parameter space.

The manuscript is organized as follows. In Section~\ref{s-mod}, we present our new model to calculate reflection spectra of thin disks in Kerr spacetimes and, in Section~\ref{s-mod2}, we show its predictions and the impact of the returning radiation in reflection spectra. In Section~\ref{s-sim}, we simulate some observations of bright Galactic black holes with \textsl{NICER} and \textsl{NuSTAR} and we present the spectral analysis of these simulated data. Discussion and conclusions are reported in Section~\ref{s-c}. In Appendix~\ref{a-a} and Appendix~\ref{a-b}, we report the key-equations to calculate the emissivity profile generated by a lamppost corona and how the returning radiation illuminates the disk, respectively. In Appendix~\ref{a-c}, we present an extension of our model in which the ionization parameter of the disk is calculated self-consistently as a function of the radial coordinate from the actual incident radiation and the disk's electron density.


\section{Description of the model}\label{s-mod}

This section describes how our model calculates reflection spectra of thin disks. In what follows, we assume that the spacetime metric is described by the Kerr solution, but our model does not exploit any specific property of the Kerr metric (e.g., the existence of the Carter constant) and we could implement another stationary, axisymmetric, and asymptotically-flat metric only by changing the metric coefficients. The accretion disk is supposed to be infinitesimally thin, on the plane perpendicular to the black hole spin. The inner edge of the disk is set at the ISCO radius, $r_{\rm in} = r_{\rm ISCO}$, and the outer edge is set at the radial coordinate $r_{\rm out} = 500$~$r_{\rm g}$. The motion of the material in the disk is Keplerian. We employ the lamppost coronal model, so the corona is a point-like source along the black hole spin axis and is only regulated by its height $h$. The emission of the corona is isotropic in its rest-frame and its spectrum is described by a power law with a high-energy cutoff, so there are two parameters: the photon index $\Gamma$ and the high-energy cutoff $E_{\rm cut}$ (which is measured in the rest-frame of the corona). With such a setup, we report below step by step the calculations of our model.

\subsection{Illumination of the disk by the corona}

To calculate how the corona illuminates the accretion disk, we modify the model {\tt blacklamp}~\citep{2019ApJ...878...91A,2022ApJ...925...51R}, which is available on GitHub\footnote{\url{https://github.com/ABHModels/blacklamp}} and Zenodo~\citep{blacklamp}. The key-equations are reported in Appendix~\ref{a-a}. First, we identify 100~reference radii on the accretion disk. We employ the following algorithm
\be\label{eq-radii}
r_i = r_{\rm in} + \left( r_{\rm out} - r_{\rm in} \right) \left(\frac{i}{i_{\rm max} - 2}\right)^3 \, ,
\ee
where $i = 0, 1, ... , i_{\rm max} - 1$ and here $i_{\rm max} = 100$. If we want to change the number of reference radii and/or the radial coordinates of the inner/outer edges of the accretion disk, we have just to change the values of, respectively, $i_{\rm max}$, $r_{\rm in}$, or $r_{\rm out}$ in Eq.~(\ref{eq-radii}).

At this point, we consider the locally-Minkowskian reference frame of the corona and we fire photons to hit the disk at the 100~reference radii in Eq.~(\ref{eq-radii}). We require that the difference between the radial coordinate at which the photon hits the disk, say $\tilde{r}_i$, and the desired reference radius, $r_i$, is less than $10^{-6}$~$r_{\rm g}$. When we meet this condition, we store the value of $\tilde{r}_i$ (which becomes our new reference radius $i$), we calculate the redshift $g_i$, and we fire one more photon to hit the disk near $\tilde{r}_i$ and calculate the energy density illuminating the disk surface at $\tilde{r}_i$ (see Appendix~\ref{a-a}).

\subsection{Non-relativistic reflection spectrum (0th order)}\label{ss-0th}

For every reference radius $\tilde{r}_i$, we calculate the non-relativistic reflection spectrum in the rest-frame of the material of the disk. Here and in the next steps, we use the code of {\tt reflionx}~\citep{2005MNRAS.358..211R}, which can calculate the non-relativistic reflection spectrum for an arbitrary input spectrum illuminating the disk. The spectrum of the radiation illuminating the reference radius $i$ is described by the photon index $\Gamma$ and the redshifted high-energy cutoff $E_{{\rm cut} , i} = g_i E_{\rm cut}$, where $g_i$ is the redshift factor between the corona and $\tilde{r}_i$. At the end, we have 100~non-relativistic reflection spectra, one for every reference radius.

A couple of comments are in order here. First, we use the code of {\tt reflionx}, not the public table of {\tt reflionx}, because the latter assumes that the incident radiation illuminating the disk has a spectrum described by a power law with a high-energy cutoff. While this would be enough at this step (non-relativistic reflection spectrum at the 0th order), in the next steps the illuminating spectrum includes a returning radiation component (see below). Second, the choice of the non-relativistic reflection model is not crucial for the purposes of this work because we do not analyze real data and we only compare theoretical predictions among different models that employ the same non-relativistic reflection calculations. We could have used the code of {\tt xillver} (but, again, not the public table of {\tt xillver}) or any other non-relativistic reflection model that can calculate the output spectrum for an arbitrary input spectrum. We do not discuss here the differences between the predictions of {\tt reflionx} and {\tt xillver}, but a work on the subject is in preparation.

\subsection{Returning radiation}\label{ss-rr}

We consider a point at the reference radius $\tilde{r}_i$ and we calculate how this point illuminates the disk. The key-equations are reported in Appendix~\ref{a-b}. We consider the locally-Minkowskian reference frame of this point and we fire photons isotropically (our photon grid is 4000$\times$4000). The photon trajectories are calculated in Boyer-Lindquist coordinates, from our reference point to the accretion disk (returning radiation), the black hole event horizon or the plunging region (radiation lost into the black hole or the plunging region), or to some large radial coordinate faraway from the disk and the black hole (radiation escaping to infinity). The trajectories of the photons hitting the disk are also the trajectories that contribute to the returning radiation of our reference point (if we consider the time reversal process). To infer the total spectrum of the returning radiation, we sum up all redshifted non-relativistic reflection spectra from those trajectories, where the non-relativistic reflection spectra at the emission points are obtained by interpolation from the spectra at the 100~reference radii calculated in Subsection~\ref{ss-0th}. We repeat these calculations for every reference radius. At the end, we have 100~spectra of the returning radiation, one for every reference radius $\tilde{r}_i$.

\subsection{Non-relativistic reflection spectrum (1st order)}\label{ss-nonrel0}

We use the code of {\tt reflionx} to calculate the non-relativistic reflection spectra for the 100~reference radii generated by the combination of the radiation from the corona and the 1st order returning radiation. At the end, we have 100 non-relativistic reflection spectra, one for every reference radius. Note that at this step it would be incorrect to use the public table of {\tt reflionx} because the spectrum illuminating the disk is not a power law with a high-energy cutoff as at the 0th order.

\subsection{Non-relativistic reflection spectrum (higher orders)}\label{ss-nonrel4}

To calculate the returning radiation at higher order, we do not to need to repeat all calculations in Subsection~\ref{ss-rr}, because the photon trajectories do not change. For every reference radius, we have to recalculate the total spectrum of the returning radiation, which is obtained by summing up all redshifted non-relativistic reflection spectra (1st order) from the emission points illuminating the points at the reference radii. At the end, we have 100~spectra of the returning radiation (2nd order), one for every reference point.

We consider two more iterations and for every reference point we obtain the spectra of the returning radiation at the 3rd and 4th orders. We do not go beyond the 4th order because we do not see any significant change in the spectra if we include higher order corrections.

\subsection{Relativistic reflection spectrum}

At this point, we have a non-relativistic reflection spectrum at 4th order for every reference radius of the accretion disk. We use the ray-tracing code {\tt blackray}\footnote{\url{https://github.com/ABHModels/blackray}}~\citep{blackray} to calculated the {\it relativistic} reflection spectrum of the whole disk as seen by a distant observer with viewing angle $\iota$. The ray-tracing code has a grid on the image plane of the distant observer. From every point of this grid, we fire a photon and we calculate the photon trajectory backward in time from the image plane of the observer to the emission point on the accretion disk. We can thus calculate the redshift associated to that point of the grid and we can determine its reflection spectrum at 4th order. At the end, we have the image of the accretion disk and every pixel of this image has its own spectrum. The total spectrum of the source is obtained by summing up the spectra of all pixels. These calculations have been already extensively discussed in the literature and the reader can find all the details, for instance, in~\citet{2017bhlt.book.....B}.

\subsection{Computational time}

To calculate a full setup, our model takes 3-4~hours using 80~threads. The most time-consuming part is the calculation of the non-relativistic reflection spectra with the {\tt reflionx} code, while the ray-tracing calculations with {\tt blacklamp} and {\tt blackray} are relatively fast.


\begin{figure}[t]
\centering
\includegraphics[width=0.95\linewidth]{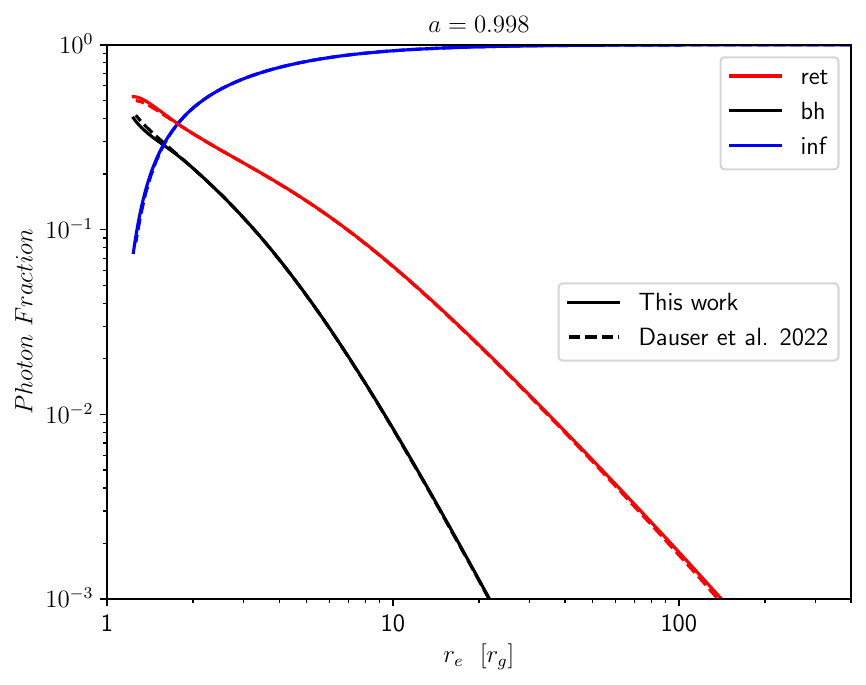}
\caption{Fraction of photons returning to the accretion disk (returning radiation, red curves), falling onto the black hole or the plunging region (black curves), and escaping to infinity (blue curves) as a function of the radial coordinate of the emission point on the disk. The solid curves are for the predictions of the model presented in this manuscript and the dashed curves are for the predictions of the model presented in \citet{2022MNRAS.514.3965D}. The photon trajectories depend on the spacetime metric and the results shown here are for a Kerr spacetime with spin parameter $a_* = 0.998$.}
\label{f-frac}
\vspace{0.5cm}
\end{figure}

\section{Predictions}\label{s-mod2}

In this section, we present the predictions of our new model. The ray-tracing calculations are compared with those in \citet{2022MNRAS.514.3965D} and we show that the predictions of the two models are in good agreement. The two models solve different equations (our model does not exploit the properties of the Kerr solution and solves equations valid for any stationary, axisymmetric, and asymptotically-flat spacetime), but the theoretical framework is almost the same. The key-feature of our new model is that it calculates the non-relativistic reflection spectrum at every point on the disk by using the actual spectrum of the incident radiation, which is the combination of a power law spectrum from the lamppost corona and of the sum of reflection spectra from different parts of the disk returning to the disk.

First, from the ray-tracing calculations described in Subsection~\ref{ss-rr}, for every point on the disk emitting reflection photons, we can evaluate the fraction of photons that returns to the disk (returning radiation), falls onto the black hole or the plunging region, and escapes to infinity. Fig.~\ref{f-frac} compares our results with those in \citet{2022MNRAS.514.3965D}. The dashed curves for \citet{2022MNRAS.514.3965D} are difficult to see in the plot because of the very good agreement between the two codes. The figure shows the calculations for a Kerr spacetime with spin parameter $a_* = 0.998$, where $r_{\rm in} = 1.237$~$r_{\rm g}$. Most of the photons emitted from the very inner part of the accretion disk ($r < 2$~$r_{\rm g}$) return to the disk or fall onto the black hole and the plunging region. Already for $r > 2$~$r_{\rm g}$, most of the photons can escape to infinity and their fraction approaches 1 as the radial coordinate increases.

\begin{figure*}
\centering
\includegraphics[width=0.95\linewidth]{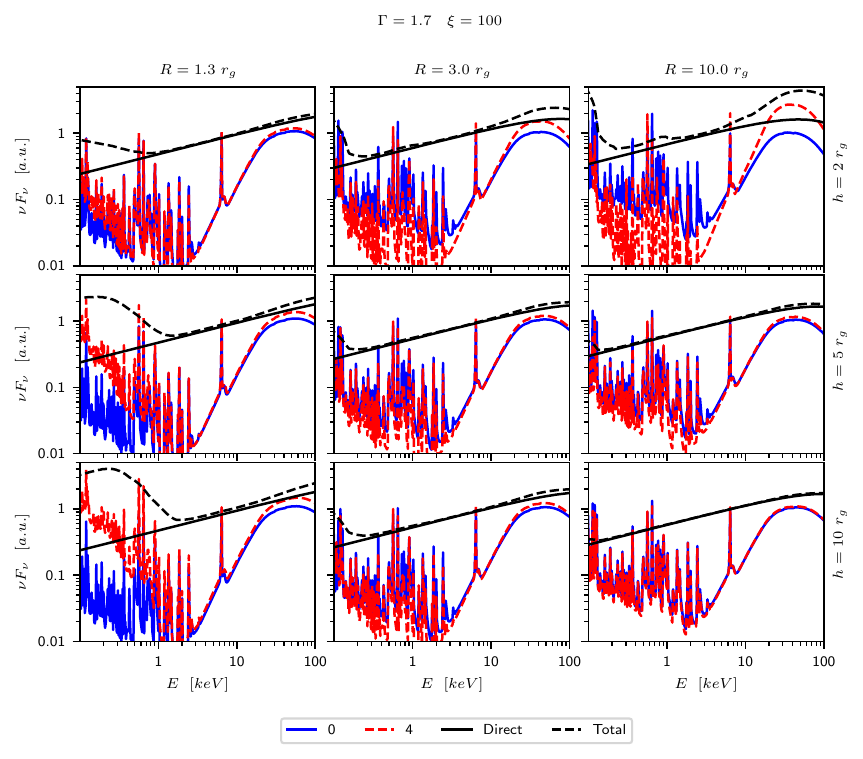}
\caption{Synthetic reflection spectra in the rest-frame of the material in the disk at different radial coordinates ($R = 1.3$, 3, and 10~$r_{\rm g}$) without returning radiation (blue solid curves, label ``0'') and with returning radiation calculated with four iterations (dashed red curves, label ``4''). In every panel, we also show the spectrum of the incident radiation from the corona (black solid curves, label ``Direct'') and the spectrum of the total incident radiation (direct radiation from the corona + returning radiation from the disk, black dashed curves, label ``Total''). The calculations assume that the spacetime is described by the Kerr metric with spin parameter $a_* = 0.998$, the spectrum of the corona at the emission point is described by a power law with photon index $\Gamma = 1.7$ and high-energy cutoff $E_{\rm cut} = 300$~keV, and the disk's ionization parameter is $\xi = 100$~erg~cm~s$^{-1}$. The y-axis shows the quantity $\nu F_\nu$ in arbitrary units, where $\nu$ is the photon frequency and $F_\nu$ is the observed flux at the frequency $\nu$. The coronal height is $h = 2$, 5, and 10~$r_{\rm g}$ for the upper, central, and lower panels, respectively. See the text for more details.}
\label{f-nonrelG17Xi2}
\end{figure*}

\begin{figure*}
\centering
\includegraphics[width=0.95\linewidth]{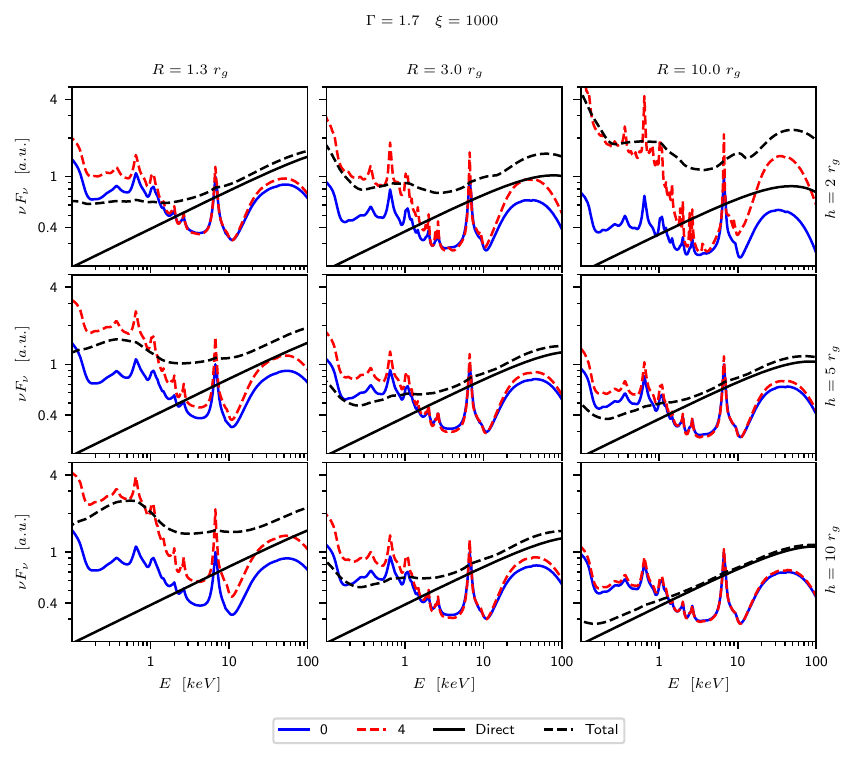}
\caption{As in Fig.~\ref{f-nonrelG17Xi2} for $\xi = 1000$~erg~cm~s$^{-1}$. See the text for more details.}
\label{f-nonrelG17Xi3}
\end{figure*}

Figs.~\ref{f-nonrelG17Xi2} and \ref{f-nonrelG17Xi3} show some non-relativistic reflection spectra as calculated in Subsections~\ref{ss-0th} and \ref{ss-nonrel4} assuming that the spectrum of the corona is described by a power law with photon index $\Gamma = 1.7$ and high-energy cutoff $E_{\rm cut} = 300$~keV, the ionization of the disk is $\xi=100$~erg~cm~s$^{-1}$ (in Fig.~\ref{f-nonrelG17Xi2}) and 1000~erg~cm~s$^{-1}$ (in Fig.~\ref{f-nonrelG17Xi3}), and the black hole spin parameter is $a_* = 0.998$. The figures show the results for three different coronal heights ($h = 2$, 5, and 10~$r_{\rm g}$) at three different radial coordinates on the accretion disk ($R = 1.3$, 3, and 10~$r_{\rm g}$). The blue solid curves are the non-relativistic reflection spectra without returning radiation (Subsection~\ref{ss-0th}) and the red dashed curves are for the non-relativistic reflection spectra with the returning radiation calculated at the 4th order (Subsection~\ref{ss-nonrel4}). In general, the iron line region is not particularly affected by the returning radiation, while larger differences between the spectra without and with returning radiation appear at lower energies (below 2-3~keV) and around the peak of the Compton hump.

Figs.~\ref{f-nonrelG17Xi2} and \ref{f-nonrelG17Xi3} show even the spectra of the incident radiation: the black solid curves are for the direct incident radiation from the corona, while the dashed black curves are for the total incident radiation illuminating the disk (i.e., direct radiation from the corona + returning radiation from the disk). As we can see from both figures, the total incident spectrum can be significantly different from a power law spectrum, so it should not be a surprise that the non-relativistic reflection spectrum can be appreciably affected by the returning radiation. In some cases, at some energies the flux of the incident radiation with returning radiation can be more than an order of magnitude higher than that without returning radiation. For $\xi=100$~erg~cm~s$^{-1}$, the contribution of the returning radiation is mainly at very low energies and around the peak of the Compton hump. For $\xi=1000$~erg~cm~s$^{-1}$, the returning radiation seems to contribute appreciably over the whole X-ray spectrum.

\begin{figure*}
\centering
\includegraphics[width=0.95\linewidth]{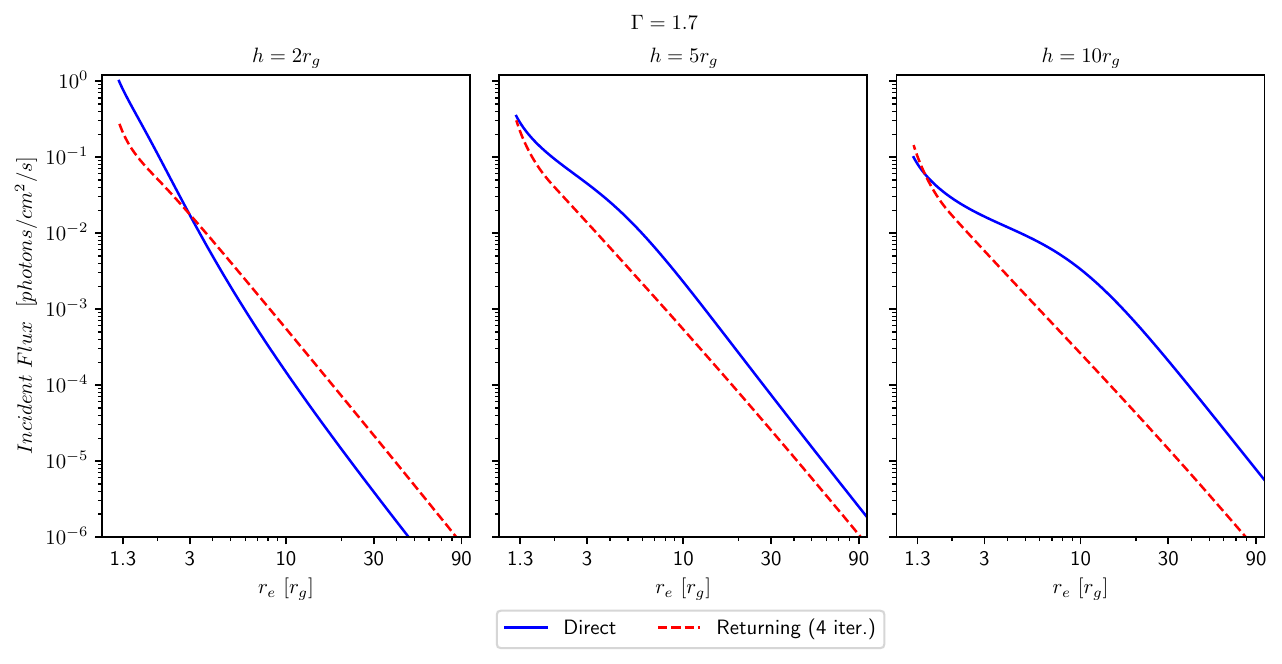}
\caption{Direct incident X-ray flux from the corona (blue solid curves) and incident X-ray flux generated by the returning radiation as calculated at the 4th order (red dashed curves) for a coronal height $h = 2$ (left panel), 5 (central panel), and 10~$r_{\rm g}$ (right panel). These results are obtained assuming that the spacetime geometry is described by the Kerr solution with spin parameter $a_* = 0.998$ and that the spectrum of the corona is described by a power law with photon index $\Gamma = 1.7$ and high-energy cutoff $E_{\rm cut} = 300$~keV.}
\label{f-profG17Xi3}
\end{figure*}

Fig.~\ref{f-profG17Xi3} shows the direct incident X-ray flux from the corona (blue solid curves) and the incident flux of the returning radiation calculated at the 4th order (red dashed curves) for a coronal height $h = 2$, 5, and 10~$r_{\rm g}$ in Kerr spacetime with $a_* = 0.998$. The spectrum of the corona is described by a power law with photon index $\Gamma = 1.7$ and high-energy cutoff $E_{\rm cut} = 300$~keV. Fig.~\ref{f-profG17Xi3} can be compared with Fig.~5 in \citet{2022MNRAS.514.3965D} and, even in this case, the two models seem to agree. If the corona is very close to the black hole, the flux of the returning radiation is lower than the direct flux from the corona in the inner part of the accretion disk and higher in the outer part (left panel in Fig.~\ref{f-profG17Xi3}). For higher values of the corona height, we can see the opposite effect, namely the incident flux of the returning radiation can exceed that from the corona in the inner part, even if only in a very small region of the accretion disk, and is lower in the outer part (right panel in Fig.~\ref{f-profG17Xi3}).

Figs.~\ref{f-nonrelG17Xi2}, \ref{f-nonrelG17Xi3}, and \ref{f-profG17Xi3} can be qualitatively explained as follows; see also \citet{2022MNRAS.514.3965D}, where such a behavior was found and discussed for the first time. If the corona is close to the black hole ($h$ is low), the effect of light bending is strong and most of the radiation illuminates the inner part of the accretion disk, while at larger radii the disk is not illuminated well. If the corona is not close to the black hole ($h$ is high), the effect of light bending is weak and there are no dramatic differences between the illumination of the disk at small and large radii. The returning radiation is mainly produced by the inner part of the accretion disk (as shown even in Fig.~\ref{f-frac}), but such a radiation does not necessarily return to the inner part of the disk. The combination of these effects leads to what we see in Figs.~\ref{f-nonrelG17Xi2}, \ref{f-nonrelG17Xi3}, and \ref{f-profG17Xi3}. 
\begin{itemize}
\item For $h = 2$~$r_{\rm g}$, the corona strongly illuminates the inner part of the accretion disk, while the contribution of the returning radiation is subdominant (in Figs.~\ref{f-nonrelG17Xi2} and \ref{f-nonrelG17Xi3}, the difference between the spectra without and with returning radiation is small for $R = 1.3$~$r_{\rm g}$ and in Fig.~\ref{f-profG17Xi3} the red dashed curve is below the blue solid curve at small radii). Since the corona directly illuminates well the inner part of the accretion disk, there are many photons that can return to the disk at large radii, where the illumination from the corona is weak and therefore the returning radiation is dominant (in Figs.~\ref{f-nonrelG17Xi2} and \ref{f-nonrelG17Xi3}, the difference between the spectra without and with returning radiation is large for $R = 10$~$r_{\rm g}$ and in Fig.~\ref{f-profG17Xi3} the red dashed curve is above the blue solid curve at large radii). 
\item For $h = 10$~$r_{\rm g}$, the corona illuminates the disk more uniformly. At small radii, the contribution of the returning radiation is relatively high because such a region is not strongly illuminated by the corona (in Figs.~\ref{f-nonrelG17Xi2} and \ref{f-nonrelG17Xi3}, the difference between the spectra without and with returning radiation is relatively large for $R = 1.3$~$r_{\rm g}$ and in Fig.~\ref{f-profG17Xi3} the red dashed curve is above the blue solid curve at small radii). At large radii, we have the opposite effect because the inner disk is not strongly illuminated and therefore there are not many photons returning to the disk and, at the same time, the outer part of the disk is already illuminated well by the corona (in Figs.~\ref{f-nonrelG17Xi2} and \ref{f-nonrelG17Xi3}, the difference between the spectra without and with returning radiation is small for $R = 10$~$r_{\rm g}$ and in Fig.~\ref{f-profG17Xi3} the red dashed curve is below the blue solid curve at large radii).
\end{itemize}

\begin{figure*}
\centering
\includegraphics[width=0.95\linewidth]{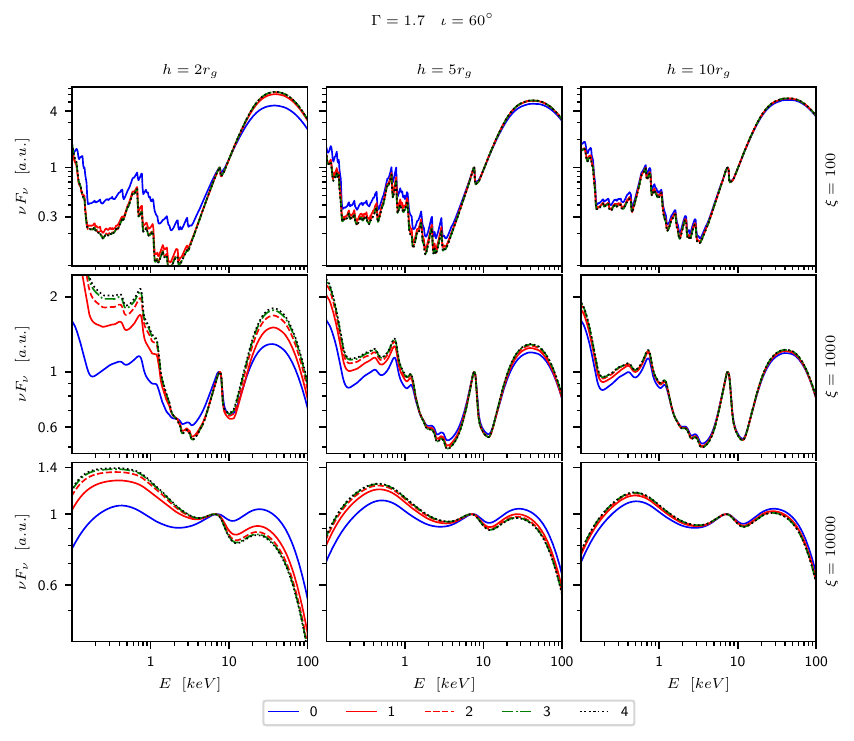}
\caption{Synthetic reflection spectra without returning radiation (blue solid curves, label ``0'') and with returning radiation calculated with one (red solid curves, label ``1''), two (red dashed curves, label ``2''), three (green dashed-dotted curves, label ``3''), and four (black dotted curves, label ``4'') iterations. We consider three different values of coronal height ($h = 2$, 5, 10~$r_{\rm g}$) and three different values of ionization parameter of the disk ($\xi=100$, 1000, and 10000~erg~cm~s$^{-1}$). The calculations assume that the spacetime is described by the Kerr metric with spin parameter $a_* = 0.998$, the spectrum of the corona at the emission point is described by a power law with photon index $\Gamma = 1.7$ and high-energy cutoff $E_{\rm cut} = 300$~keV, and the inclination angle of the disk with respect to the line of sight of the observer is $\iota = 60^\circ$. The y-axis shows the quantity $\nu F_\nu$ in arbitrary units ($\nu$ is the photon frequency and $F_\nu$ is the observed flux at the frequency $\nu$) and normalized to have the peak of the iron K$\alpha$ line at 1 in order to facilitate the comparison of the shapes of the spectra. See the text for more details.}
\label{f-G17i60}
\end{figure*}

\begin{figure*}
\centering
\includegraphics[width=0.95\linewidth]{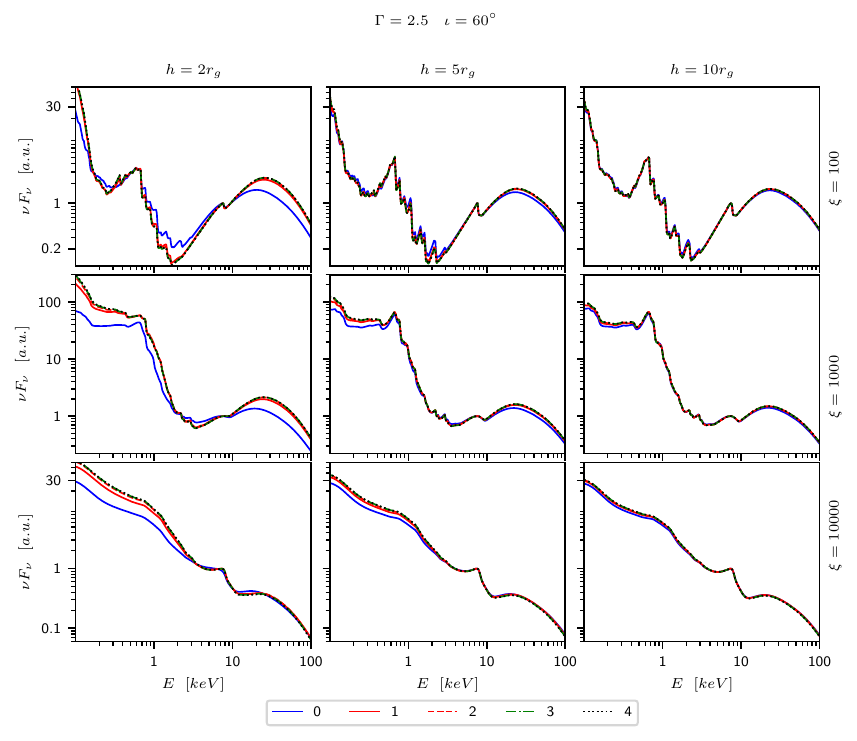}
\caption{As in Fig.~\ref{f-G17i60} for $\Gamma = 2.5$ and $\iota = 60^\circ$.}
\label{f-G25i60}
\vspace{0.78cm}
\end{figure*}

\begin{figure*}
\centering
\includegraphics[width=0.95\linewidth]{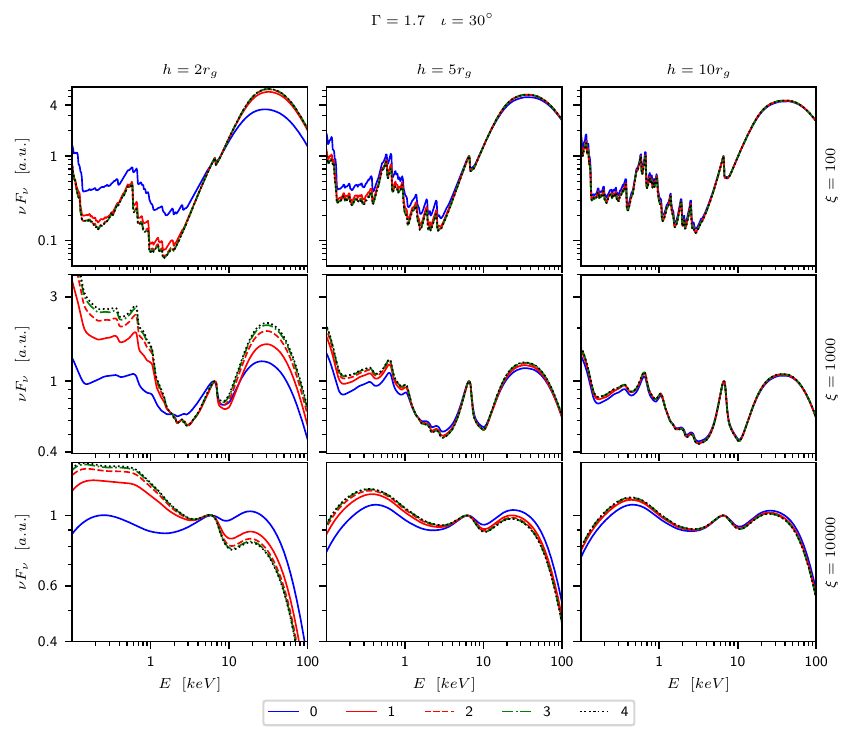}
\caption{As in Fig.~\ref{f-G17i60} for $\Gamma = 1.7$ and $\iota = 30^\circ$.}
\label{f-G17i30}
\vspace{1.0cm}
\end{figure*}

\begin{figure*}
\centering
\includegraphics[width=0.95\linewidth]{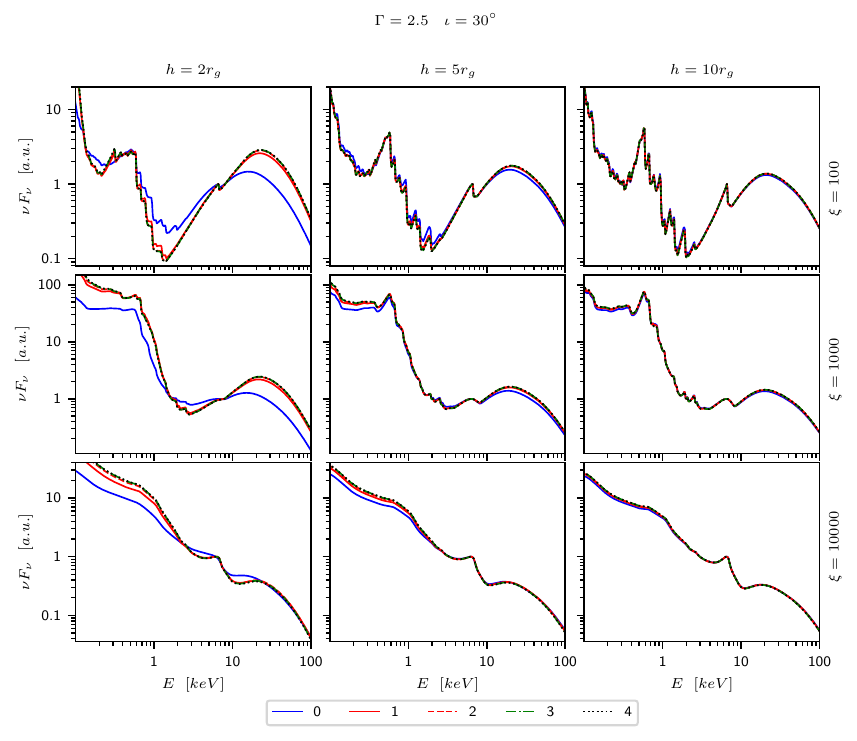}
\caption{As in Fig.~\ref{f-G17i60} for $\Gamma = 2.5$ and $\iota = 30^\circ$.}
\label{f-G25i30}
\vspace{1.0cm}
\end{figure*}

Relativistic reflection spectra without returning radiation and with returning radiation calculated at the 1st, 2nd, 3rd, and 4th orders are shown in Fig.~\ref{f-G17i60}, Fig.~\ref{f-G25i60}, Fig.~\ref{f-G17i30}, and Fig.~\ref{f-G25i30}. Every figure shows the results for three different coronal heights ($h = 2$, 5, 10~$r_{\rm g}$) and three different disk's ionization parameters ($\xi=100$, 1000, and 10000~erg~cm~s$^{-1}$). The inclination angle of the disk with respect to the line of sight of the observer is $\iota = 60^\circ$ in Fig.~\ref{f-G17i60} and Fig.~\ref{f-G25i60} and $\iota = 30^\circ$ in Fig.~\ref{f-G17i30} and Fig.~\ref{f-G25i30}. The photon index of the coronal spectrum is $\Gamma = 1.7$ (Fig.~\ref{f-G17i60} and Fig.~\ref{f-G17i30}) and 2.5 (Fig.~\ref{f-G25i60} and Fig.~\ref{f-G25i30}). The high-energy cutoff is always $E_{\rm cut} = 300$~keV. We note that all spectra are normalized to have the peak of the iron K$\alpha$ line at $\nu F_\nu = 1$ in order to compare better the shapes of the spectra (without such a normalization, the flux of the spectra without returning radiation would be lower than the flux of the spectra with returning radiation, as the latter increases the incident flux illuminating the disk). The height of the corona is the most important parameter to regulate the difference between the spectra without and with returning radiation. This is understandable, because photons returning to the disk are mainly emitted from the very inner part of the accretion disk, and the emissivity of the inner part of the accretion disk increases as the coronal height decreases.

\begin{table*}[tbh]
\renewcommand{\arraystretch}{1.5}
\centering
\begin{tabular}{c c c c c c c c}
\hline\hline
\hspace{0.3cm} Simulation \hspace{0.3cm} & \hspace{0.3cm}  $a_*$ \hspace{0.3cm} & \hspace{0.3cm} $h$ $[r_{\rm g}]$ \hspace{0.3cm} & \hspace{0.3cm} $i$ [deg] \hspace{0.3cm} & \hspace{0.3cm} $\Gamma$ \hspace{0.3cm} & \hspace{0.3cm} $E_{\rm cut}$ [keV] \hspace{0.3cm} & \hspace{0.3cm} $\xi$ [${\rm erg}\cdot{\rm cm}\cdot{\rm s}^{-1}$] \hspace{0.3cm} & \hspace{0.3cm} $A_{\rm Fe}$ \hspace{0.3cm} \\
\hline
A1 & 0.998 & 2 & $60$ & 1.7 & 300 & 100 & 1 \\
A2 & 0.998 & 2 & $60$ & 1.7 & 300 & 1000 & 1 \\
A3 & 0.998 & 2 & $60$ & 1.7 & 300 & 10000 & 1 \\
B1 & 0.998 & 5 & $60$ & 1.7 & 300 & 100 & 1 \\
B2 & 0.998 & 5 & $60$ & 1.7 & 300 & 1000 & 1 \\
B3 & 0.998 & 5 & $60$ & 1.7 & 300 & 10000 & 1 \\
\hline\hline
\end{tabular}
\vspace{0.2cm}
\caption{\rm Summary of the input parameters of the simulations considered in this work with $a_* = 0.998$. In simulations A, the coronal height is $h = 2~r_{\rm g}$, while in simulations B we have $h = 5~r_{\rm g}$. The disk ionization parameter is $\xi = 100$, 1000, and 10000~erg~cm~s$^{-1}$ for, respectively, simulations 1, 2, and 3. The values of the other parameters do not change among different simulations. \label{t-sim}}
\vspace{0.0cm}
         \renewcommand\arraystretch{1.5}
	\centering
	\vspace{0.5cm}
	\begin{tabular}{lcccccc}
		\hline\hline
		            & \hspace{0.5cm} A1 \hspace{0.5cm} & \hspace{0.5cm} A2 \hspace{0.5cm} & \hspace{0.5cm} A3 \hspace{0.5cm} & \hspace{0.5cm} B1 \hspace{0.5cm} & \hspace{0.5cm} B2 \hspace{0.5cm} & \hspace{0.5cm} B3 \hspace{0.5cm} \\
		\hline
		$h~[r_{\rm g}]$ & 
  $16.2_{-2.2}^{+2.5}$ & 
  $2.02_{-(B)}^{+0.04}$ & 
  $21_{-8}^{+6}$ & 
  $6.0_{-0.6}^{+0.6}$ & 
  $5.5_{-0.6}^{+0.6}$ & 
  $2.850_{-0.020}^{+4.00}$ \\ 
		$a_*$ & 
 $0.10_{-(B)}^{+0.08}$ & 
  $0.9827_{-0.0023}^{+0.0013}$ & 
  $0.998_{-(B)}^{+(B)}$ & 
  $0.10_{-(B)}^{+0.10}$ & 
  $0.10_{-(B)}^{+0.20}$ & 
  $0.10_{-(B)}^{+0.7}$ \\ 
		$i$ [deg] &
  $68_{-4}^{+5}$ & 
  $61.60_{-0.14}^{+0.14}$ & 
  $42.8_{-2.6}^{+3.0}$ & 
  $59_{-0.4}^{+0.3}$ & 
  $60.0_{-0.5}^{+0.5}$ & 
  $52.8_{-1.2}^{+1.2}$ \\ 
		$\Gamma$ & 
  $1.7128_{-0.0023}^{+0.0021}$ & 
  $1.82554_{-0.0020}^{+0.0021}$ & 
  $1.700_{-0.011}^{+0.012}$ & 
  $1.700_{-0.005}^{+0.005}$ & 
  $1.700_{-0.005}^{+0.005}$ & 
  $1.700_{-0.006}^{+0.003}$ \\
        $E_{\rm cut}$ [keV]  &
  $435_{-40}^{+25}$ & 
  $500_{-2}^{+(B)}$ & 
  $390_{-20}^{+20}$ & 
  $340_{-17}^{+19}$ & 
  $350_{-12}^{+14}$ & 
  $500_{-(B)}^{+(B)}$ \\
		$\xi$ [${\rm erg}\cdot{\rm cm}\cdot{\rm s}^{-1}$] &
  $10.0_{-(B)}^{+1.0}$ & 
  $390.0_{-1.0}^{+1.0}$ & 
  $9000.0_{-1.0}^{+1.0}$ & 
  $72_{-1.1}^{+1.1}$ & 
  $700.0_{-1.0}^{+1.0}$ & 
  $8700_{-7}^{+11}$ \\
		$A_{\rm Fe}$ &
  $0.620_{-0.013}^{+0.013}$ & 
  $1.200_{-0.024}^{+0.027}$ & 
  $0.95_{-0.06}^{+0.06}$ & 
  $0.940_{-0.016}^{+0.016}$ & 
  $0.985_{-0.014}^{+0.030}$ & 
  $0.90_{-0.07}^{+0.05}$ \\ 
		\hline
		\hline
		$\chi^2/\nu $ &
  $1359.36/437$ &
  $1796.57/431$ &
  $438.48/420$ &
  $615.83/437$ &
  $698.62/429$ &
  $416.29/423$ \\ 
		& 
  $=3.11066$ & 
  $=4.16837$ &
  $=1.04400$ &
  $=1.40922$ &
  $=1.62848$ &
  $=0.98413$ \\		\hline\hline
	\end{tabular}
	\vspace{0.2cm}
	\caption{\rm Best-fit values of simulations A1--A3 and B1--B3 with returning radiation. The reported uncertainties correspond to the 90\% confidence level for one relevant parameter ($\Delta\chi^2=2.71$). $(B)$ means that the 90\% confidence level reaches the boundary of the parameter or the best-fit value is stuck at the boundary of the parameter. The values of the input parameters are reported in Tab.~\ref{t-sim}.} \label{t-bfv}
\end{table*}

\begin{figure*}
\centering
\includegraphics[width=0.47\linewidth]{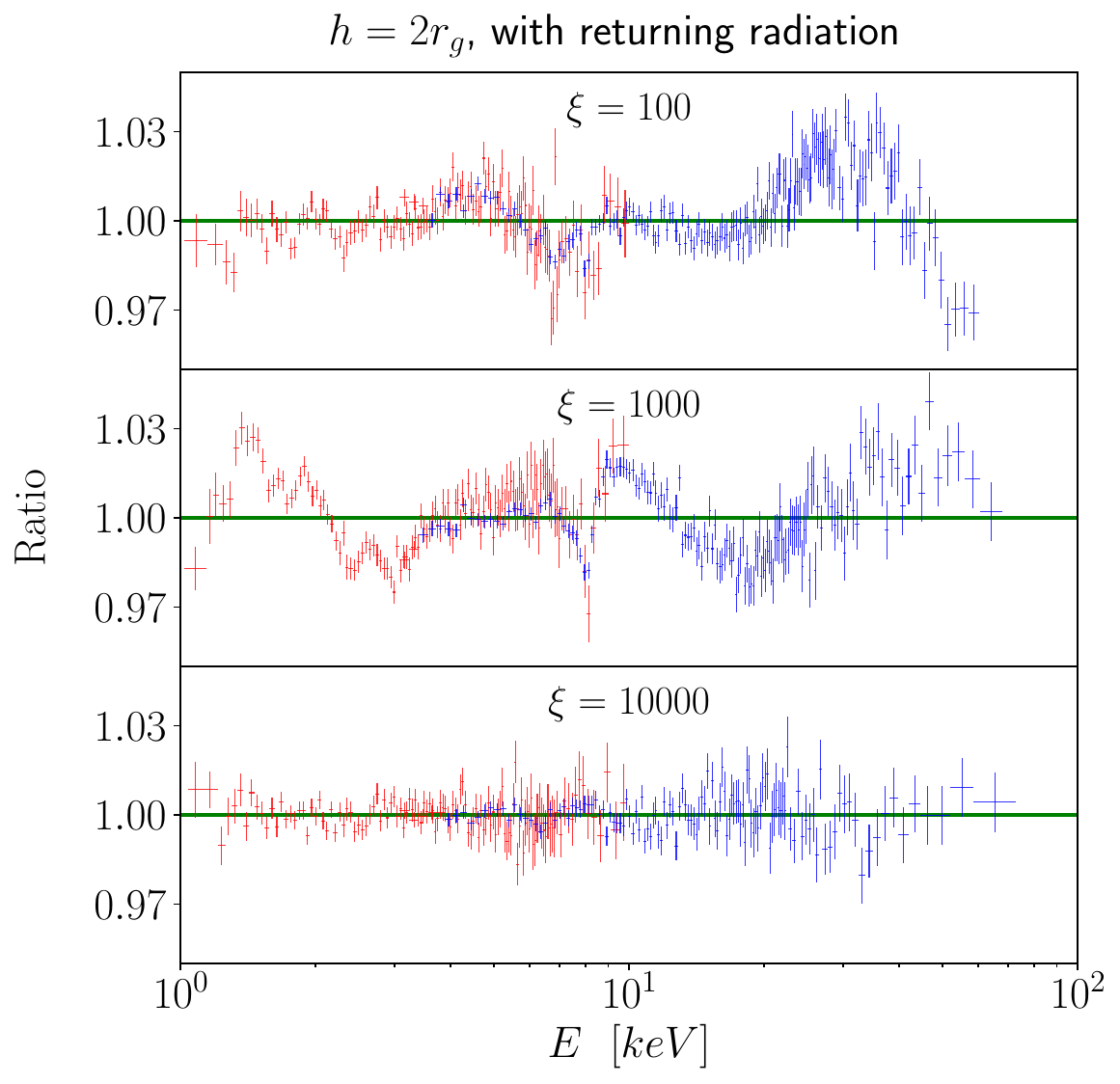}
\hspace{0.3cm}
\includegraphics[width=0.47\linewidth]{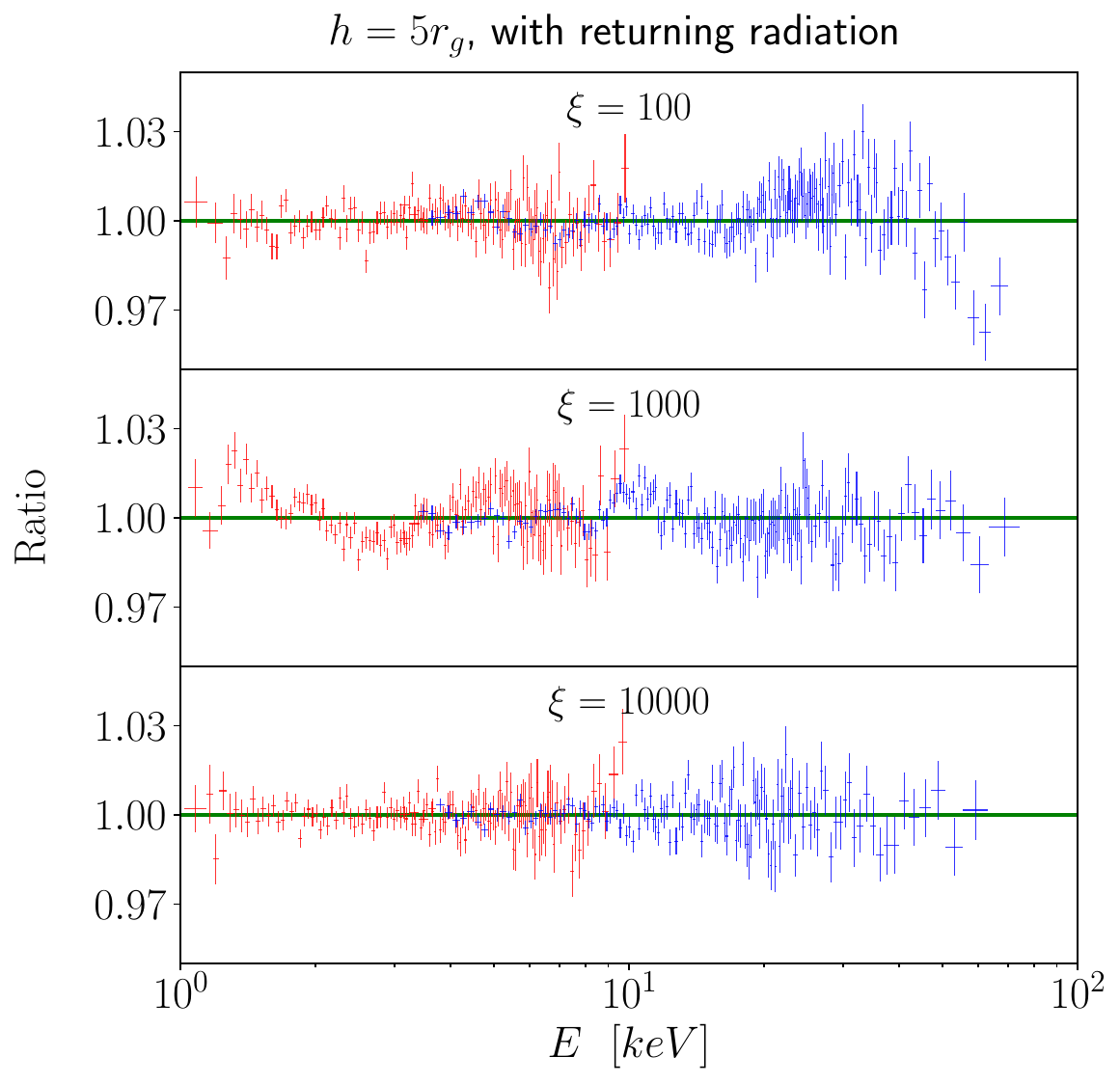}
\caption{Data to best-fit model ratios for simulations A1--A3 ($h = 2~r_{\rm g}$ and $a_* = 0.998$) and B1--B3 ($h = 5~r_{\rm g}$ and $a_* = 0.998$) with returning radiation. The red and blue colors represent, respectively, the \textsl{NICER} and \textsl{NuSTAR} data. $\xi$ in units erg~cm~s$^{-1}$.}
\label{f-sims-a}
\end{figure*}


\section{Simulations and fits}\label{s-sim}

In the latest version of the {\tt relxill} package, the effect of the returning radiation is taken into account only in the emissivity profile, while the non-relativistic reflection spectra are still those calculated assuming that the incident radiation has a power law spectrum with a high-energy cutoff~\citep{2022MNRAS.514.3965D}. Such an approximation is motivated by the fact that the actual incident spectrum, which is the result of the combination of the direct spectrum from the corona and the returning radiation, has a shape rather close to a power law; see Fig.~7 in \citet{2022MNRAS.514.3965D}. The key-parameters are the black hole spin, the coronal height, and the disk's ionization parameter. The black hole spin and the coronal height regulate the intensity of the returning radiation. The ionization parameter regulates the shape of the reflection spectrum: if the ionization is low, there are many features in the reflection spectrum, but these features are washed out when the value of the ionization parameter becomes very high (and there are no features when the disk is fully ionized).

To quantify the capability of the {\tt relxill} package to analyze real sources, we simulate some observations of bright Galactic black holes. We assume that the energy flux of the source is $10^{-8}$~erg~cm$^{-2}$~s$^{-1}$ in the 1--10~keV energy range. In XSPEC language, the model of the simulations is

\vspace{0.3cm}

\noindent {\tt tbabs$\times$(cutoffpl + reflection)} \, ,

\vspace{0.3cm}

\noindent where {\tt tbabs} takes the Galactic absorption into account~\citep{2000ApJ...542..914W}, {\tt cutoffpl} describes the Comptonized spectrum of the corona, and {\tt reflection} is the reflection spectrum calculated by our model. The input parameters are summarized in Tab.~\ref{t-sim}. We consider two coronal heights ($h = 2$ and 5~$r_{\rm g}$) and three possible disk's ionization parameters ($\xi=100$, 1000, and 10000~erg~cm~s$^{-1}$). The other input parameters are the same for all simulations. The black hole spin parameter is set to $a_* = 0.998$ to maximize the impact of the returning radiation.

For the same input parameters, we run a first simulation without returning radiation and a second simulation with returning radiation calculated at the 4th order. We assume simultaneous observations of \textsl{NICER} and \textsl{NuSTAR} and we consider an exposure time of 5~ks for \textsl{NICER} and of 30~ks for \textsl{NuSTAR} (for simplicity, we run the simulations assuming 60~ks with one detector rather than 30~ks with two detectors).

We fit the simulated data with the model

\vspace{0.3cm}

\noindent {\tt tbabs$\times$(cutoffpl + relxilllp\_reflionx)} \, ,

\vspace{0.3cm}
 
\noindent where {\tt relxilllp\_reflionx} is the lamppost flavor of the latest version of the {\tt relxill} package~\citep{2022MNRAS.514.3965D}, in which the returning radiation can be turned off and on, and has been modified to read the {\tt reflionx} table instead of the {\tt xillver} one. In this way, we use the same model as in the simulations to calculate non-relativistic reflection spectra and we include the flux correction factor discussed in Appendix~B in \citet{2022MNRAS.514.3965D}. If we used {\tt relconvlp$\times$reflionx}, the fitting model would not include such a flux correction factor, while such an effect is automatically included by construction in the calculations of our new model.

In {\tt relxilllp\_reflionx} we turn the returning radiation off to fit the simulations without returning radiation and we turn it on to fit the simulations with returning radiation calculated at the 4th order. For the simulations without returning radiation, we recover the correct input parameters and the fits are good, in the sense that we do not see unresolved features in the plots of the ratios between the simulated data and the best-fit models. This is just a healthy check for our new model and we do not report here the results because the latter are perfectly consistent with our expectations of good fits and correct estimate of the model parameters.

For the simulations with returning radiation, the quality of the fits is not always good and some parameters are significantly underestimated or overestimated. The results of these fits with returning radiation are summarized in Tab.~\ref{t-bfv} and the residuals are shown in Fig.~\ref{f-sims-a}. In the data to the best-fit model ratios, there are large residuals for simulations A1 and A2, there are still some residuals for simulations B1 and B2, while there are no clear residuals for simulations A3 and B3 with a high value of the ionization parameter $\xi$. The discussion of these fits is postponed to the next section.


\section{Discussion and conclusions}\label{s-c}

For a qualitative discussion of the impact of the returning radiation on observed reflection spectra, we can refer to Fig.~\ref{f-G17i60}, Fig.~\ref{f-G25i60}, Fig.~\ref{f-G17i30}, and Fig.~\ref{f-G25i30}. After fixing the black hole spin parameter to $a_* = 0.998$, the two key-parameters are the coronal height $h$ and the disk's ionization parameter $\xi$, while the impact of the values of the photon index $\Gamma$ and of the inclination angle of the disk $\iota$ is weaker. Within the lamppost setup, the coronal height $h$ is the parameter regulating the intensity of the returning radiation simply because it determines the emissivity profile of the direct radiation from the corona: as shown in Fig.~\ref{f-frac} (where the three curves depend only weakly on the black hole spin parameter except for the fact $a_*$ determines the inner edge of the disk, $r_{\rm in}$) only a significant fraction of photons emitted at small radii returns to the disk, and therefore the effect of the returning radiation increases/decreases when the emission at small radii increases/decreases. The ionization parameter $\xi$ determines scattering probabilities and energy lines, but its impact on the spectra is more complicated. For $\xi = 100$~erg~cm~s$^{-1}$, the returning radiation decreases the relative intensity of the reflection features in the soft X-ray band and increases the intensity of the Compton hump, while we see the opposite trend for $\xi = 10000$~erg~cm~s$^{-1}$. On the other hand, for $\xi = 1000$~erg~cm~s$^{-1}$, it seems like the intensity of the reflection features in the soft X-ray band and the Compton hump are both increased with respect to the iron line.

For a quantitative discussion of the impact of the returning radiation, we have the simulations and fits presented in Section~\ref{s-sim}. We remind the reader that our simulations are for a Kerr spacetime with spin parameter $a_* = 0.998$, which represents the situation in which the observed spectra are more strongly affected by the returning radiation. Moreover, we are assuming a static corona: in the case of a rotating corona, the impact of the returning radiation would be weaker~\citep{1997MNRAS.288L..11D,2008MNRAS.386..759N}. First, we note that the quality of the fits with $h = 2$~$r_{\rm g}$ and a low or moderate value of the ionization parameter $\xi$ is bad: the reduced $\chi^2$ is unacceptably high and there are large residuals in the ratio plots. For simulation~A1 with $\xi = 100$~erg~cm~s$^{-1}$, the coronal height is clearly overestimated and the ionization parameter $\xi$, the iron abundance $A_{\rm Fe}$, and the black hole spin parameter $a_*$ are underestimated. For simulation~A2 with $\xi = 1000$~erg~cm~s$^{-1}$, the fit recovers relatively well the coronal height and the black hole spin parameter, even if the quality of the fit is still bad. If we increase the coronal height to $h = 5$~$r_{\rm g}$, the quality of the fits improves: in simulations B1 and B2, the reduced $\chi^2$ is around 1.5 and in the ratio plots we do not see the large residuals of the fits of simulations A1 and A2. The value of the coronal height is recovered quite well, while the estimate of the spin is not consistent with the spin parameter of the simulations (see Tab.~\ref{t-bfv}). In the two cases with a high ionization parameter (A3 and B3), the quality of the fits is good, but mainly because the reflection features are weak. In the fit of simulation A3, the coronal height is clearly overestimated, the spin parameter is unconstrained, and the inclination angle of the disk is underestimated. In the case of the fit of B3, the spin parameter is poorly unconstrained and we do not recover the correct value at the 90\% confidence level. So for high values of the ionization parameter we may not recover the correct input parameter without a clear signal in the residuals.

\begin{figure*}
\centering
\includegraphics[width=0.95\linewidth]{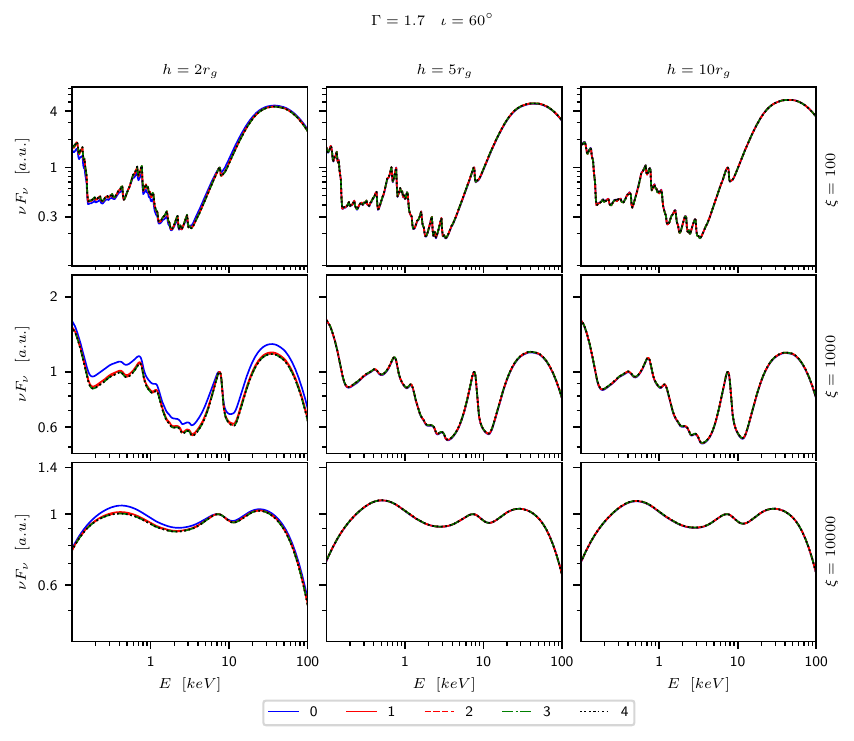}
\caption{As in Fig.~\ref{f-G17i60} but calculating the reflection spectrum assuming that the spectrum of the radiation illuminating the disk is described by a power law with a high-energy cutoff as in \citet{2022MNRAS.514.3965D}.}
\label{f-simple}
\vspace{0.8cm}
\end{figure*}

We would like to stress that the main feature of our model is that we calculate the reflection spectrum at every radial coordinate of the accretion disk taking into account the actual spectrum illuminating the disk, which is a combination of a power law spectrum from the corona and a reflection spectrum returning to the disk. The four iterations described in Subsection~\ref{ss-nonrel4} are necessary to find the final reflection spectrum, while higher order effects have a negligible impact on the emissivity profile of the disk. To prove such a statement, we rerun our model assuming that the incident spectrum illuminating the disk is always a power law with a high-energy cutoff, and we take the returning radiation into account only in the intensity of the radiation, which is the approximation adopted in \citet{2022MNRAS.514.3965D}. The results for $\Gamma = 1.7$ and $\iota = 60^\circ$ are shown in Fig.~\ref{f-simple}, where we clearly see that the impact of the returning radiation is much weaker than in Fig.~\ref{f-G17i60} and the 1st order calculations are enough, in agreement with the conclusion reported in \citet{2022MNRAS.514.3965D}. We note that we have checked our emissivity profile with such an approximation with the emissivity profile of {\tt relxilllp\_reflionx} with returning radiation and the two profiles agree.

\begin{figure*}
\centering
\includegraphics[width=0.47\linewidth]{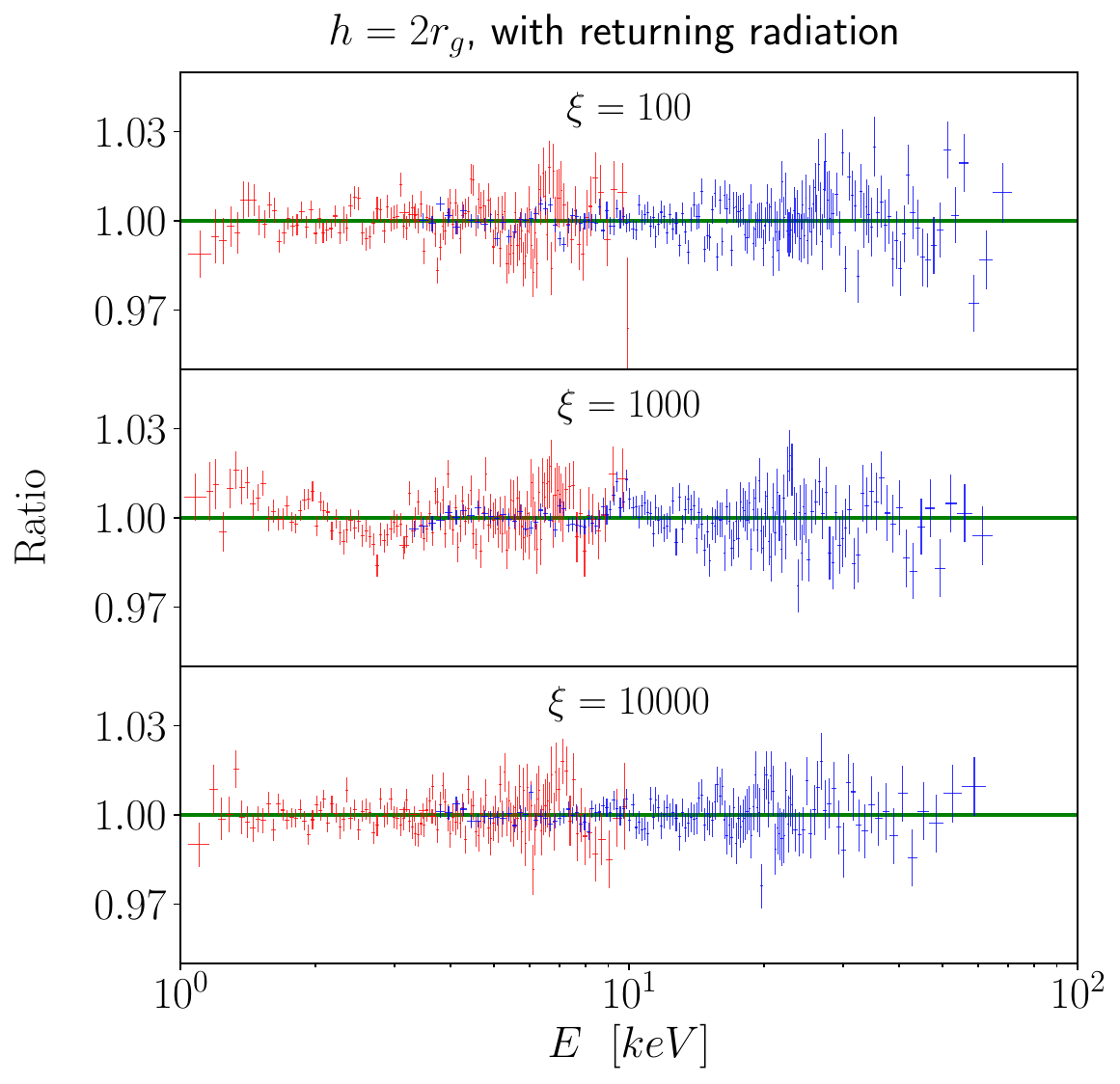}
\hspace{0.3cm}
\includegraphics[width=0.47\linewidth]{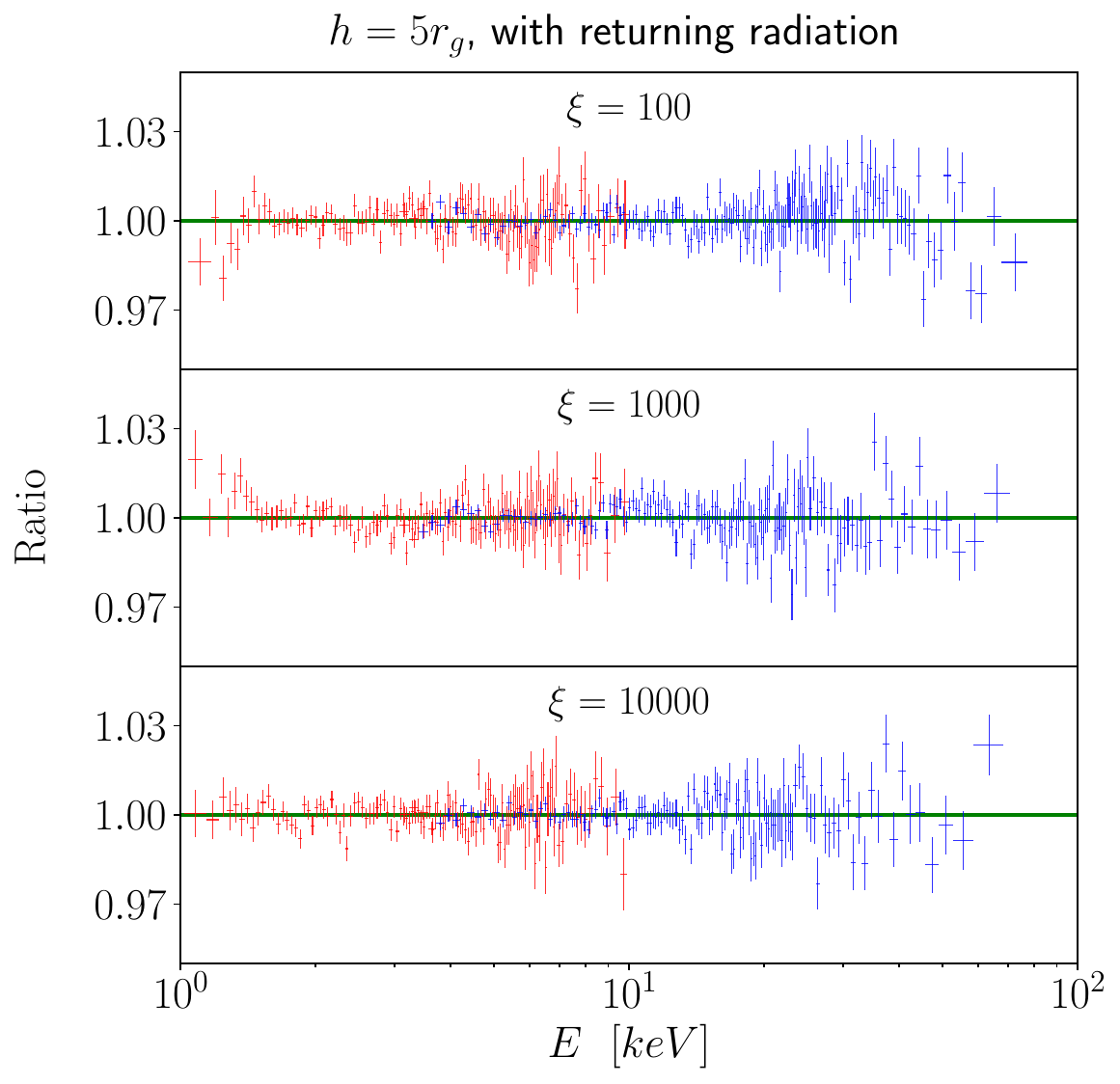}
\caption{Data to best-fit model ratios for simulations C1--C3 ($h = 2~r_{\rm g}$ and $a_* = 0.9$) and D1--D3 ($h = 5~r_{\rm g}$ and $a_* = 0.9$) with returning radiation. The red and blue colors represent, respectively, the \textsl{NICER} and \textsl{NuSTAR} data. $\xi$ in units erg~cm~s$^{-1}$.}
\label{f-sims-b}
\end{figure*}

In the previous section, all simulations assumed $a_* = 0.998$ to maximize the impact of the returning radiation. If we decrease the value of the black hole spin parameter, the inner edge of the accretion disk (which is supposed to be at the ISCO radius in this work) moves to larger radii. Since it is mainly the radiation emitted at very small radii to return to the disk (see Fig.~\ref{f-frac}), we automatically decrease the impact of the returning radiation on the observed spectrum. To be more quantitative, we run another cycle of simulations with $a_* = 0.9$. The input parameters are summarized in Tab.~\ref{t-simbis} and they are exactly the same as in Tab.~\ref{t-sim} except for the value of the black hole spin parameter. We assume the same properties for the sources and for the observations of \textsl{NICER} and \textsl{NuSTAR}. We fit the simulated data with {\tt tbabs$\times$(cutoffpl + relxilllp\_reflionx)} and the ratio plots are shown in Fig.~\ref{f-sims-b} (left panels for $h=2$~$r_{\rm g}$, right panels for $h=5$~$r_{\rm g}$). Tab.~\ref{t-bfvbis} shows the best-fit values of the fits with returning radiation. The quality of the fit is definitively better than the fits for the simulations with $a_* = 0.998$: the reduced $\chi^2$ is relatively close to 1 even in simulations C1 and C2, in Fig.~\ref{f-sims-b} we do not see the large residuals of Fig.~\ref{f-sims-a}, and we substantially recover all the input parameters (with some exceptions).

\begin{table*}[tbh]
\renewcommand{\arraystretch}{1.5}
\centering
\begin{tabular}{c c c c c c c c}
\hline\hline
\hspace{0.3cm} Simulation \hspace{0.3cm} & \hspace{0.3cm}  $a_*$ \hspace{0.3cm} & \hspace{0.3cm} $h$ $[r_{\rm g}]$ \hspace{0.3cm} & \hspace{0.3cm} $i$ [deg] \hspace{0.3cm} & \hspace{0.3cm} $\Gamma$ \hspace{0.3cm} & \hspace{0.3cm} $E_{\rm cut}$ [keV] \hspace{0.3cm} & \hspace{0.3cm} $\xi$ [${\rm erg}\cdot{\rm cm}\cdot{\rm s}^{-1}$] \hspace{0.3cm} & \hspace{0.3cm} $A_{\rm Fe}$ \hspace{0.3cm} \\
\hline
C1 & 0.9 & 2 & $60$ & 1.7 & 300 & 100 & 1 \\
C2 & 0.9 & 2 & $60$ & 1.7 & 300 & 1000 & 1 \\
C3 & 0.9 & 2 & $60$ & 1.7 & 300 & 10000 & 1 \\
D1 & 0.9 & 5 & $60$ & 1.7 & 300 & 100 & 1 \\
D2 & 0.9 & 5 & $60$ & 1.7 & 300 & 1000 & 1 \\
D3 & 0.9 & 5 & $60$ & 1.7 & 300 & 10000 & 1 \\
\hline\hline
\end{tabular}
\vspace{0.2cm}
\caption{\rm Summary of the input parameters of the simulations with $a_* = 0.9$. In simulations C, the coronal height is $h = 2~r_{\rm g}$, while in simulations D we have $h = 5~r_{\rm g}$. The disk ionization parameter is $\xi = 100$, 1000, and 10000~erg~cm~s$^{-1}$ for, respectively, simulations 1, 2, and 3. The values of the other parameters do not change among different simulations. \label{t-simbis}}
\vspace{0.0cm}
         \renewcommand\arraystretch{1.5}
	\centering
	\vspace{0.5cm}
	\begin{tabular}{lcccccc}
		\hline\hline
		            & \hspace{0.5cm} C1 \hspace{0.5cm} & \hspace{0.5cm} C2 \hspace{0.5cm} & \hspace{0.5cm} C3 \hspace{0.5cm} & \hspace{0.5cm} D1 \hspace{0.5cm} & \hspace{0.5cm} D2 \hspace{0.5cm} & \hspace{0.5cm} D3 \hspace{0.5cm} \\
		\hline
		$h~[r_{\rm g}]$ & 
  $10.6_{-0.7}^{+0.7}$ & 
  $2.23_{-0.14}^{+0.30}$ & 
  $6.1_{-3.0}^{+3.0}$ & 
  $5.8_{-0.9}^{+2.0}$ & 
  $5.7_{-0.6}^{+0.5}$ & 
  $2.0_{-(B)}^{+8.0}$ \\ 
		$a_*$ & 
 $0.92_{-0.09}^{+(B)}$ & 
  $0.900_{-0.006}^{+0.005}$ & 
  $0.76_{-(B)}^{+(B)}$ & 
  $0.20_{-(B)}^{+0.7}$ & 
  $0.86_{-0.08}^{+0.04}$ & 
  $0.94_{-0.2}^{+0.03}$ \\ 
		$i$ [deg] &
  $60.8_{-0.6}^{+0.6}$ & 
  $59.5_{-0.40}^{+0.22}$ &
  $54.0_{-1.1}^{+1.2}$ & 
  $59.0_{-0.4}^{+0.5}$ & 
  $60.0_{-0.3}^{+0.4}$ & 
  $60.0_{-0.8}^{+0.7}$ \\ 
		$\Gamma$ & 
  $1.700_{-0.006}^{+0.005}$ & 
  $1.700_{-0.005}^{+0.007}$ & 
  $1.700_{-0.003}^{+0.004}$ & 
  $1.700_{-0.005}^{+0.005}$ & 
  $1.700_{-0.004}^{+0.006}$ & 
  $1.675_{-0.004}^{+0.006}$ \\ 
        $E_{\rm cut}$ [keV]  &
  $220_{-9}^{+8}$ & 
  $285_{-6}^{+6}$ & 
  $216_{-31}^{+6}$ & 
  $320_{-16}^{+20}$ &
  $290_{-7}^{+6}$ & 
  $378_{-12}^{+18}$ \\ 
		$\xi$ [${\rm erg}\cdot{\rm cm}\cdot{\rm s}^{-1}$] &
  $71.0_{-1.0}^{+1.0}$ & 
  $820_{-2}^{+3}$ & 
  $8550_{-3}^{+4}$ & 
  $81_{-1.0}^{+1.0}$ &  
  $840.0_{-1.2}^{+1.2}$ & 
  $9000_{-10}^{+60}$ \\ 
		$A_{\rm Fe}$ &
  $0.900_{-0.013}^{+0.007}$ & 
  $0.995_{-0.007}^{+0.032}$ & 
  $1.12_{-0.10}^{+0.11}$ & 
  $0.968_{-0.015}^{+0.016}$ & 
  $0.995_{-0.013}^{+0.038}$ & 
  $1.00_{-0.10}^{+0.14}$ \\ 
		\hline
		\hline
		$\chi^2/\nu $ &
  $581.07/435$ &
  $605.95/429$ &
  $383.69/422$ & 
  $489.77/435$ & 
  $449.41/428$ & 
  $475.65/424$ \\ 
		& 
  $=1.33579$ & 
  $=1.41247$ &
  $=0.90921$ &
  $=1.12590$ &
  $=1.05002$ & 
  $=1.12181$ \\		\hline\hline
	\end{tabular}
	\vspace{0.2cm}
	\caption{\rm Best-fit values of simulations C1--C3 and D1--D3 with returning radiation. The reported uncertainties correspond to the 90\% confidence level for one relevant parameter ($\Delta\chi^2=2.71$). $(B)$ means that the 90\% confidence level reaches the boundary of the parameter or the best-fit value is stuck at the boundary of the parameter. The values of the input parameters are reported in Tab.~\ref{t-simbis}.} \label{t-bfvbis}
\end{table*}

\begin{figure*}
\centering
\includegraphics[width=0.47\linewidth]{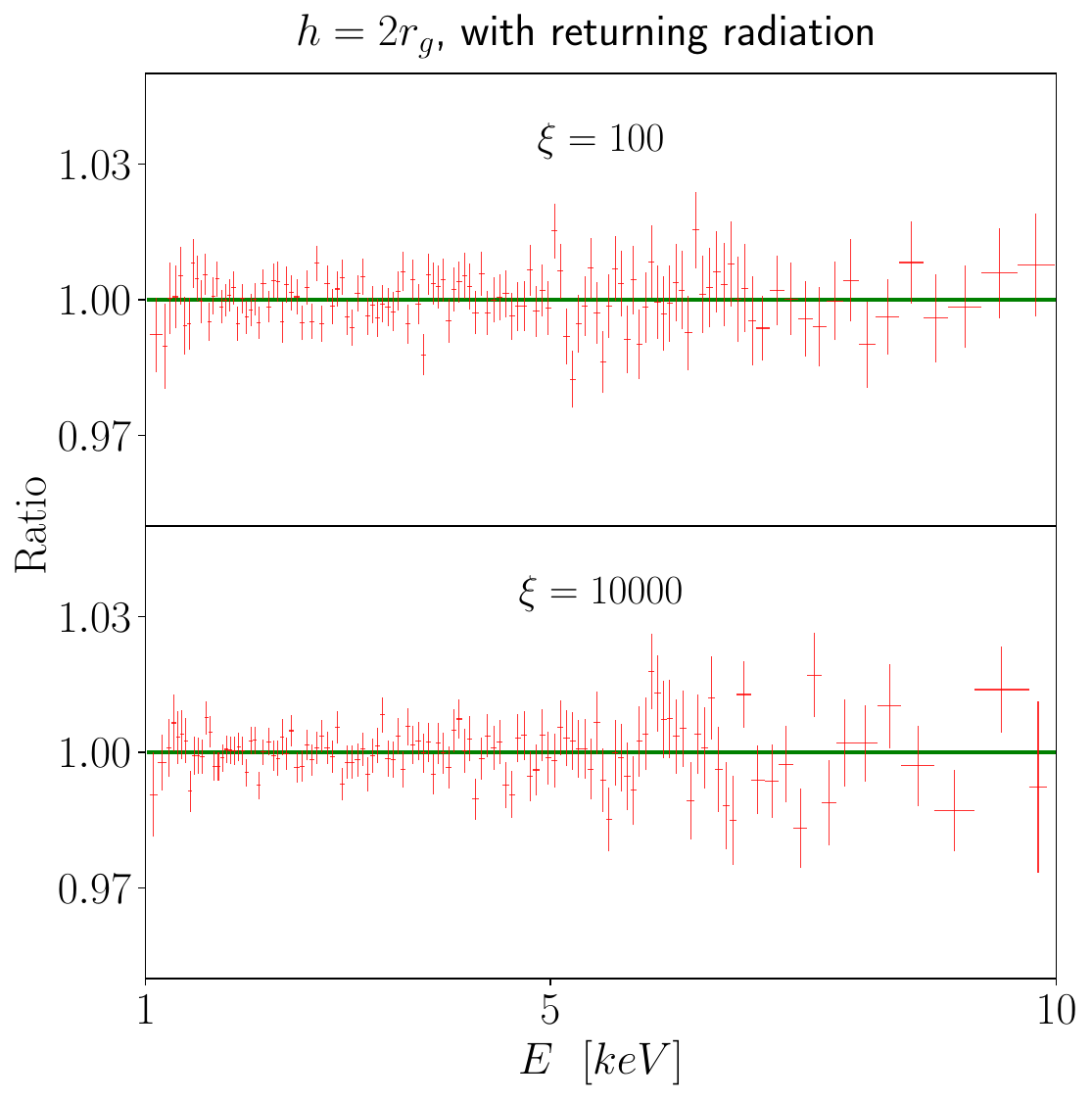}
\hspace{0.3cm}
\includegraphics[width=0.47\linewidth]{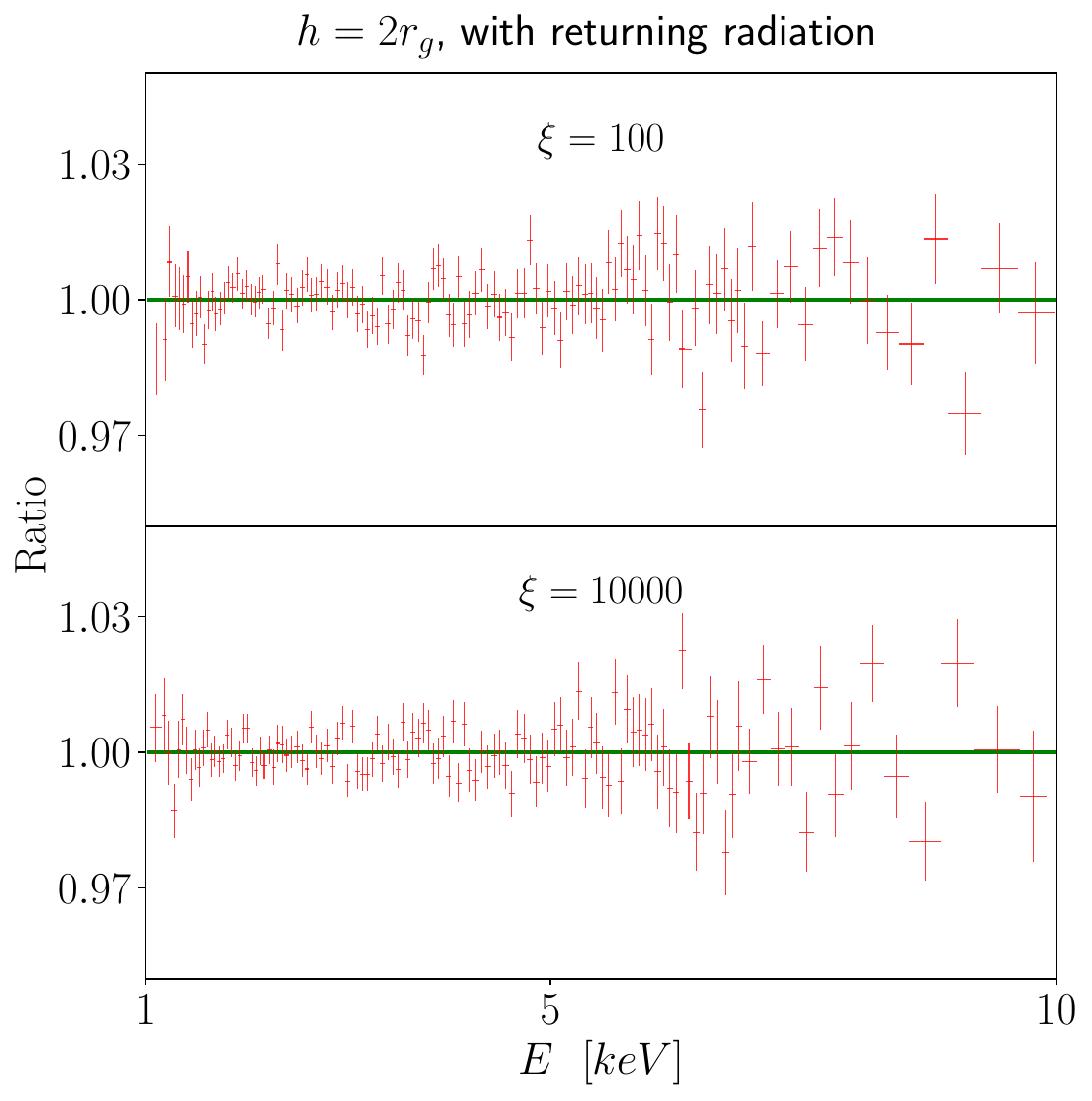}
\caption{{Left panel: Data to best-fit model ratios for simulations A1 ($a_* = 0.998$ and $\xi = 100$~erg~cm~s$^{-1}$) and A3 ($a_* = 0.998$ and $\xi = 10000$~erg~cm~s$^{-1}$) with returning radiation assuming to have only \textsl{NICER} data. Right panel: Data to best-fit model ratios for simulations C1 ($a_* = 0.9$ and $\xi = 100$~erg~cm~s$^{-1}$) and C3 ($a_* = 0.9$ and $\xi = 10000$~erg~cm~s$^{-1}$) with returning radiation assuming to have only \textsl{NICER} data.}}
\label{f-sim-nicer}
\end{figure*}

In \citet{2021ApJ...910...49R}, it was shown that, when we fit synthetic spectra that include the returning radiation with a reflection model that does not include the returning radiation, the results of the fits can depend significantly on the considered energy band. That study was limited to neutral disks. With the model presented in the current paper, we can check if such a conclusion remains true for a more realistic ionized disk. We thus repeat some fits assuming that we have only \textsl{NICER} observations. We refit simulations A1 ($h = 2$~$r_{\rm g}$, $a_* = 0.998$, and $\xi = 100$~erg~cm~s$^{-1}$), A3 ($h = 2$~$r_{\rm g}$, $a_* = 0.998$, and $\xi = 10000$~erg~cm~s$^{-1}$), C1 ($h = 2$~$r_{\rm g}$, $a_* = 0.9$, and $\xi = 100$~erg~cm~s$^{-1}$), and C3 ($h = 2$~$r_{\rm g}$, $a_* = 0.9$, and $\xi = 10000$~erg~cm~s$^{-1}$). The results of these new fits are reported in Tab.~\ref{t-sims-nicer} and Fig.~\ref{f-sim-nicer}. The results are quite different from the fits of the simulated observations with \textsl{NICER}+\textsl{NuSTAR}. We often recover the correct input parameters and there are no very large residuals in the ratio plots. In particular, we recover the correct black hole spin parameter in A1, A3, and C1, while $a_*$ cannot be constrained in C3. The values of the coronal height $h$ and of the inclination angle $i$ are recovered correctly (but in the fit of simulation A3 the value of $h$ is poorly constrained). We thus recover the conclusions of \citet{2021ApJ...910...49R} for ionized disks: if we have only data in the soft X-ray band, the impact of the returning radiation is weak.

\begin{table*}[t]
         \renewcommand\arraystretch{1.5}
	\centering
	\vspace{0.5cm}
	\begin{tabular}{lcccccc}
		\hline\hline
		            & \hspace{0.5cm} A1  \hspace{0.5cm} & \hspace{0.5cm} A3 \hspace{0.5cm} & \hspace{0.5cm} C1  \hspace{0.5cm} & \hspace{0.5cm} C3 \hspace{0.5cm} \\
		\hline
		$h~[r_{\rm g}]$ & 
  $1.95_{-(B)}^{+0.26}$ & 
  $22_{-20}^{+4}$ & 
  $2.0_{-(B)}^{+0.03}$ & 
  $3.560_{-1.4}^{+1.6}$ &  \\ 
		$a_*$ & 
 $0.988_{-0.009}^{+(B)}$ & 
  $0.997_{-0.004}^{+(B)}$ & 
  $0.890_{-0.004}^{+0.020}$ & 
  $0.972_{-(B)}^{+(B)}$ & \\ 
		$i$ [deg] &
  $60_{-2}^{+2}$ & 
  $58_{-3}^{+4}$ & 
  $60_{-0.5}^{+0.6}$ & 
  $58_{-4}^{+4}$ &  \\ 
		$\Gamma$ & 
  $1.713_{-0.015}^{+0.007}$ & 
  $1.700_{-0.019}^{+0.018}$ & 
  $1.700_{-0.010}^{+0.010}$ & 
  $1.659_{-0.017}^{+0.016}$ &  \\
        $E_{\rm cut}$ [keV]  &
  $460_{-300}^{+(B)}$ & 
  $300_{-21}^{+145}$ & 
  $412_{-200}^{+32}$ & 
  $143_{-40}^{+50}$ & \\
		$\xi$ [${\rm erg}\cdot{\rm cm}\cdot{\rm s}^{-1}$] &
  $25.0_{-2}^{+3}$ & 
  $10000_{-40}^{+10}$ & 
  $63_{-2}^{+6}$ & 
  $8000_{-20}^{+20}$ &  \\
		$A_{\rm Fe}$ &
  $0.83_{-0.21}^{+0.09}$ & 
  $1.00_{-0.13}^{+0.20}$ & 
  $0.90_{-0.06}^{+0.06}$ & 
  $0.73_{-0.11}^{+0.11}$ &  \\ 
		\hline
		$\chi^2/\nu $ &
  $135.21/141$ &
  $135.23/141$ &
  $169.93/141$ &
  $158.38/141$ &\\ 
		& 
  $=0.95893$ & 
  $=0.95907$ &
  $=1.20517$ &
  $=1.12326$ & \\	
  \hline\hline
	\end{tabular}
	\vspace{0.2cm}
	\caption{\rm {Best-fit values for simulations A1, A3, C1, and C3 with returning radiation, assuming that we have only \textsl{NICER} data. The reported uncertainties correspond to the 90\% confidence level for one relevant parameter ($\Delta\chi^2=2.71$). $(B)$ means that the 90\% confidence level reaches the boundary of the parameter or the best-fit value is stuck at the boundary of the parameter. The values of the input parameters are reported in Tab.~\ref{t-sim} and Tab.~\ref{t-simbis}.}} \label{t-sims-nicer}
\end{table*}

We conclude this section with two considerations. First, in the model presented in this paper, the returning radiation is only represented by reflection photons. This is a good approximation for accretion disks of supermassive black holes (where the thermal spectrum of the disk is in the UV band) and of Galactic black holes in the hard state and at low luminosities (where the thermal component is negligible with respect to the continuum from the corona and the reflection spectrum). In general, even the thermal radiation could contribute to the returning radiation, and it is also possible that the returning radiation of the thermal spectrum acts as a corona in sources in the soft state~\citep[see, for instance,][]{2021ApJ...909..146C}. In a future paper, we plan to present an extension of our model that includes the calculations of the thermal spectrum and where the returning radiation is generated by both the reflection and thermal components.

Second, our new model can calculate reflection spectra with returning radiation calculated with the correct spectrum illuminating the disk at every radial coordinate, but it is too slow to be used to analyze real X-ray spectra. With the current architecture of {\tt relxill} and the other reflection models, there is no straightforward way to include the returning radiation as calculated in the present work. This is because in all these models reflection spectra are obtained by calculating the integral \citep[see, for instance,][]{2017bhlt.book.....B}
\be
F_{\rm obs} (\nu_{\rm obs}) = \int_{r_{\rm in}}^{r_{\rm out}} \int_0^1 \frac{\pi r_e g^2}{\sqrt{g^* (1 - g^*)}} \; f \; I_e \; dg^* \, dr_e \, , \hspace{0.5cm}
\ee
where $g$ is the redshift factor, $g^*$ is the relative redshift factor, $f$ is the transfer function, and $I_e$ is the specific intensity of the radiation in the rest-frame of the material in the disk. The transfer functions (encoding all details about photon trajectories and particle motion in the disk) are tabulated in a FITS file for a grid of black hole spins and viewing angles. Non-relativistic reflection spectra and lamppost emissivity profiles, which determine $I_e$, are tabulated in other two FITS files. These quantities are tabulated in FITS files to be able to calculate many spectra quickly during the data analysis process, when we need to scan the parameter space and find the best-fit values of all free parameters. Especially the calculation of the non-relativistic reflection spectra, which requires to solve radiative transfer equations, is very time-consuming. In order to include the calculations of the returning radiation with the actual incident spectrum, we would need to add to the model a very heavy FITS file, beyond the capability of the RAM of current normal computers. Our plan for the near future is to develop a reflection model based on a library of reflection spectra. While the construction of such a library is certainly time-consuming, it can be done on a large computer cluster. The model can then use this library to calculate quickly many spectra and analyze real data.


{\bf Acknowledgments --} 
This work was supported by the Natural Science Foundation of Shanghai, Grant No.~22ZR1403400, the National Natural Science Foundation of China (NSFC), Grant No.~12250610185, 11973019, and 12261131497, and the Shanghai Municipal Education Commission, Grant No.~2019-01-07-00-07-E00035.
T.M. acknowledges also the support from the China Scholarship Council (CSC), Grant No.~2022GXZ005433.
S.R. acknowledges also the support from the China Postdoctoral Science Foundation, Grant No.~2022M720035, and the Teach@T\"ubingen Fellowship.
J.J. acknowledges support from Leverhulme Trust, Isaac Newton Trust and St Edmund's College, University of Cambridge.


\appendix

\section{Appendix~A: Calculation of the emissivity profile produced by a lamppost corona}\label{a-a}

The calculation of the emissivity profile produced by a lamppost corona is already discussed in the literature. Here we simply report the main steps and the key-equations.

In the locally-Minkowski reference frame of the corona, the initial conditions of the photon 4-momentum can be written as
\be
k_{(\alpha)} = ( -E_0, \; 0, \; E_0 \sin \delta, \; -E_0 \cos \delta) \, ,
\ee
where $\delta$ is the angle between the black hole spin axis and the direction of the photon 4-momentum. In Boyer-Lindquist coordinates, the initial conditions of the photon 4-momentum become
\be
k_{\mu} = (-E, \; k_{(z)} \sqrt{g_{rr}}, \; k_{(y)} \sqrt{g_{\theta \theta}}, \; 0) \, ,
\ee
where $E$ is the conserved photon energy and can be inferred from the condition $k^\mu k_\mu = 0$. The redshift factor of the photon emitted from the corona and hitting the disk at the radial coordinate $r$ is
\be
g= \frac{\nu_{c}}{\nu_{d}} = \frac{k^{c}_{\mu} u^{\mu}_{c}}{k^{d}_{\mu} u^{\mu}_{d}} \, ,
\ee
where $k^{d}_{\mu}$ is the photon 4-momentum on the disk, $u^{\mu}_{d}$ is the 4-velocity of the material in the accretion disk, $k^{c}_{\mu}$ is the photon 4-momentum at the corona, and $u^{\mu}_{c}$ is the 4-velocity of the corona. The intensity of the radiation illuminating the disk is 
\be
I \propto \dv{\Omega_{o}}{A_{o}} = \frac{1}{\bar{\gamma} \sqrt{g_{\phi\phi} g_{rr}}}\left|\dv{\cos\delta}{r}\right| \, ,
\ee
where ${\rm d}\Omega_{o}$ is the solid angle in the source rest-frame, ${\rm d}A_{o}$ is the surface area in the reference frame of the disk patch, $\bar{\gamma}$ is the Lorentz factor of the material in the accretion disk as measured in the locally-non-rotating reference frames
\be\label{eq-a-gamma}
\bar{\gamma}=\left[1+\frac{\left(\Omega_{\mathrm{K}} g_{\phi \phi} + g_{t \phi}\right)^2}{g_{t t} g_{\phi \phi}-g_{t \phi}^2}\right]^{-1 / 2} \, ,
\ee
and $\Omega_K$ is the angular velocity of the material in the accretion disk.
Last, the energy density illuminating the disk surface (= emissivity) at the reference radius $\tilde{r}_i$ is given by
\be
\epsilon_i = \frac{g_i^\Gamma}{\bar{\gamma} \sqrt{g_{\phi\phi} g_{rr}}}\left|\dv{\cos\delta}{r}\right|_{r = \tilde{r}_i} \, .
\ee

\section{Appendix~B: Calculation of the returning radiation}\label{a-b}

The calculation of the returning radiation is described in Subsection~\ref{ss-rr}. For every reference radius, we consider a point on the disk and we fire photons isotropically. In the locally-Minkowski reference frame of the emission point, the initial conditions of the photon 4-momentum can be written as
\be
k_{(\alpha)} = ( -E_0, \; E_0 \sin \theta_e \cos \phi_e, \; E_0 \sin \theta_e \sin \phi_e, \; E_0 \cos \theta_e) \, ,
\ee
where $\theta_e$ and $\phi_e$ are the emitting angles. In the locally-non-rotating reference frames, the photon 4-momentum $k_{(\alpha)}$ becomes
\be
\bar{k}_{(\alpha)} = ( \bar{\gamma} (k_{(t)} + \bar{u}^{(\phi)} k_{(y)}), \; 
 k_{(x)}, \; \bar{\gamma} (k_{(y)} + \bar{u}^{(\phi)} k_{(t)}), \; k_{(z)}) \, ,
\ee
where $\bar{\gamma}$ is still the Lorentz factor of the material in the accretion disk given in Eq.~(\ref{eq-a-gamma}) and $u^{(\phi)}$ is the $\phi$ component of the 4-velocity of the material in the disk as measured in the locally-non-rotating reference frames. The photon 4-momentum in Boyer-Lindquist coordinates are
\be
k_{\mu} = (-E, \;  \bar{k}_{(x)} \sqrt{g_{rr}}, \; -\bar{k}_{(z)} \sqrt{g_{\theta \theta}}, \; \bar{k}_{(y)} \sqrt{g_{\phi \phi}}) \, ,
\ee
where $E$ is the conserved photon energy and can be inferred from the condition $k^\mu k_\mu = 0$.
The total X-ray flux of the returning radiation at the $k$-th order illuminating the reference point at the radial coordinate $r$ is
\be
F_{\rm ret}^k(r, E) = \sum_{j} g_{j}^{3} \; I_e^k(r_j, \frac{E}{g_j}) \; \Delta \Omega_j \, \cos\theta_j \, ,
\ee
where the summation is over all the photons that return to the disk. The photon $j$ returns to the disk from the radial coordinate $r_j$ and with redshift factor $g_j$. $\theta_j$ is the incident angle of the photon $j$ at the radial coordinate $r$. The total X-ray flux illuminating the reference point at the radial coordinate $r$ at the $k$-th order is the sum of the direct X-ray flux from the corona and X-ray flux from the returning radiation at the $k$-th order
\be
F^k(r, E) = F_{\rm lp}(r, E) + F_{\rm ret}^k(r, E) \, .
\ee

\section{Appendix~C: Relativistic reflection spectra with a self-consistent ionization parameter}\label{a-c}

In the calculations presented so far in this manuscript, the disk electron density is always assumed to be $n_e = n_0 = 10^{15}$~cm$^{-3}$ at any radial coordinate. The value of the ionization parameter $\xi$ changes (we have presented most results for three cases, $\xi = 100$, 1000, and 10000~erg~cm~s$^{-1}$), but is also constant over the whole disk. However, for a real accretion disk we can expect that the electron density has a non-trivial radial profile, so $n_e = n_e (r)$. At the same time, the ionization parameter is defined as
\be\label{eq-xi}
\xi = \frac{4 \pi F_X}{n_e} \, ,
\ee 
where $F_X$ is the total X-ray flux illuminating the disk (direct X-ray flux from the corona + returning radiation). In general, both $F_X$ and $n_e$ are functions of the radial coordinate $r$ and therefore even $\xi$ should be a function of $r$. For the lamppost model, we can calculate $F_X(r)$. If we assume some radial profile for the disk electron density, we can determine $\xi = \xi(r)$ and use the correct ionization parameter at every point of the disk in the calculation of reflection spectra\footnote{We note, however, that the analysis of specific spectra has shown that models with non-trivial electron density and ionization parameter profiles can improve only marginally the quality of the fits, without significant differences in the estimates of the values of the parameters of those systems~\citep{2021ApJ...923..175A,2022MNRAS.517.5721M}.}. We note that, depending on the radial profiles of $L_X$ and $n_e$, $\xi(r)$ may not be a monotonic function of the radial coordinate $r$, and the maximum value of $\xi$, say $\xi_{\rm max}$, may not be at $r_{\rm in}$.

\begin{figure*}
\centering
\includegraphics[width=0.95\linewidth]{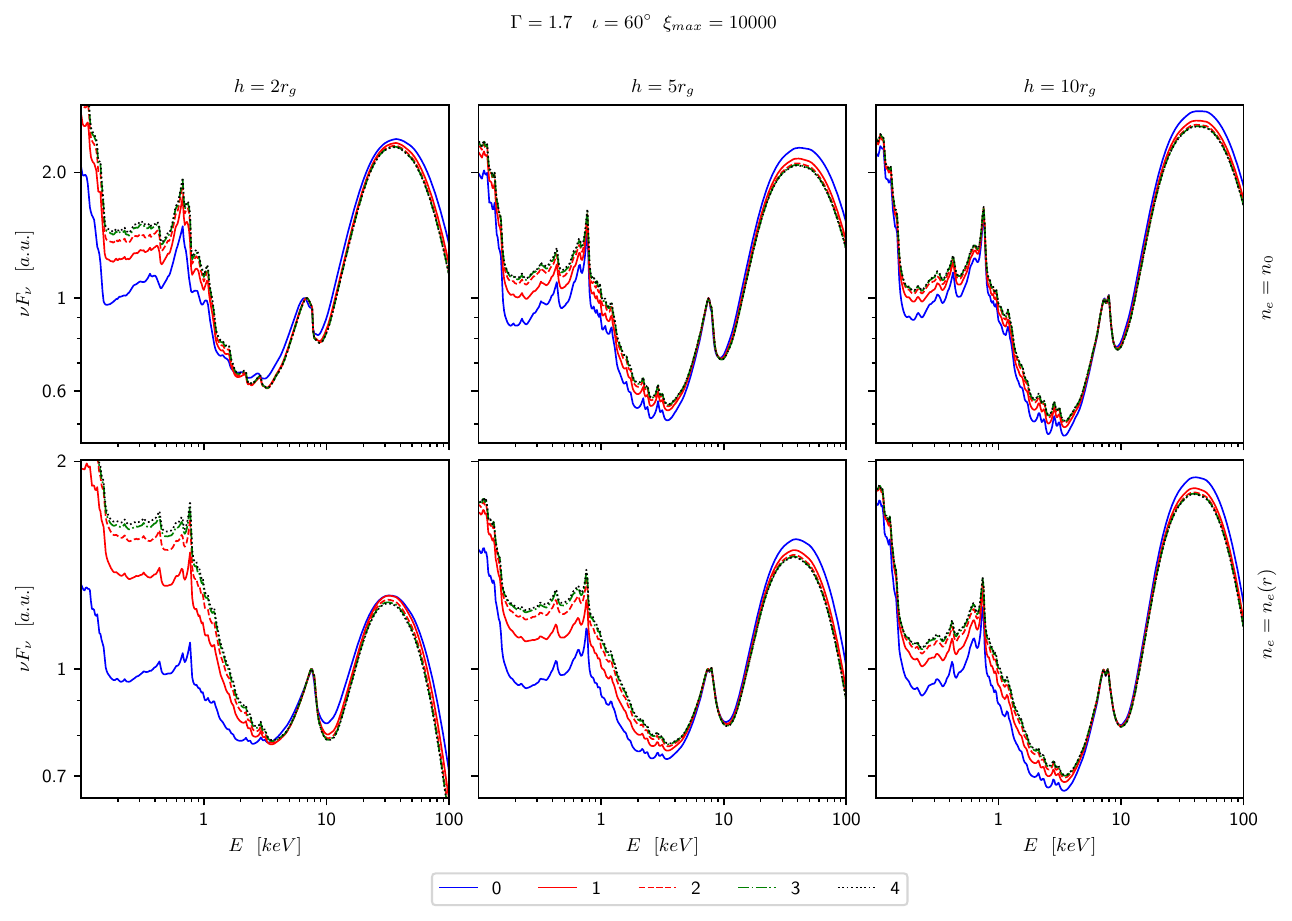}
\caption{As in Fig.~\ref{f-G17i60} with the ionization parameter $\xi$ calculated with Eq.~(\ref{eq-xi}). In the upper panels, we still assume the same electron density over the whole disk, $n_e = n_0 = 10^{15}$~cm$^{-3}$. In the lower panels, we assume the electron density profile of the Shakura-Sunyaev model, $n_e(r) \propto r^{3/2} [1 - (r_{\rm in}/r)^{1/2}]^{-2}$. See the text for more details.}
\label{f-VarIon}
\end{figure*}

\begin{figure*}
\centering
\includegraphics[width=0.95\linewidth]{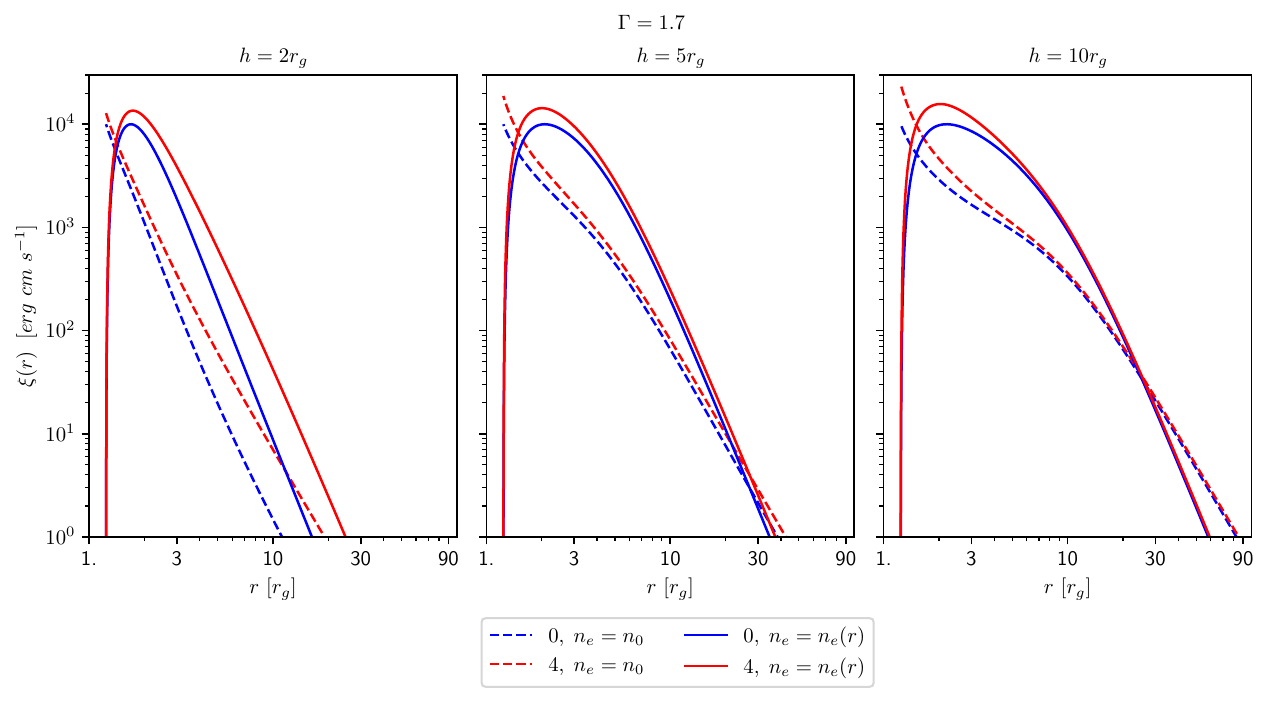}
\caption{Radial ionization profiles for the models shown in Fig.~\ref{f-VarIon}. We have two models of the disk electron density profile: $n_e = n_0 = 10^{15}$~cm$^{-3}$ (dashed curves) and the Shakura-Sunyaev profile $n_e(r) \propto r^{3/2} [1 - (r_{\rm in}/r)^{1/2}]^{-2}$ (solid curves). The blue curves are the ionization parameter profiles without returning radiation and the red curves are the profiles calculated taking the returning radiation into account.}
\label{f-IonGrad}
\end{figure*}

It is straightforward to implement a non-trivial disk electron density profile in our new model and calculate the ionization parameter at every radial coordinate of the disk with Eq.~(\ref{eq-xi}). Fig.~\ref{f-VarIon} shows synthetic reflection spectra without and with returning radiation for two electron density profiles. In the upper panels, we still assume a trivial profile, where the electron density is constant over the disk. In the lower panels, we consider the density profile of the Shakura-Sunyaev model~\citep{1973A&A....24..337S}
\be
n_e(r) \propto \frac{r^{3/2}}{[1 - (r_{\rm in}/r)^{1/2}]^2} \, ,
\ee
where the minimum of $n_e(r)$ is set to $n_0 = 10^{15}$~cm$^{-3}$. The height of the corona determines the profile of $F_X$, but not its normalization, which depends on the coronal luminosity. In our model, we regulate the coronal luminosity by changing the maximum value of the disk's ionization parameter, $\xi_{\rm max}$. In Fig.~\ref{f-VarIon}, we assume $\xi_{\rm max} = 10000$~erg~cm~s$^{-1}$. The calculations are done assuming that the spectrum of the corona is described by a power law with photon index $\Gamma=1.7$ and high-energy cutoff $E_{\rm cut} = 300$~keV, and that the inclination angle of the disk with respect to the line of sight of the observer is $\iota = 60^\circ$. Fig.~\ref{f-IonGrad} shows the ionization parameter profiles for the two choices of the electron density profile of Fig.~\ref{f-VarIon} and the coronal heights $h=2$, 5, and 10~$r_{\rm g}$. The exact choice of the electron density profile has a strong impact on the profile of $\xi$, while the impact of the returning radiation is much weaker. The returning radiation slightly increases the value of the ionization parameter resulting from the illumination of the corona.

\end{document}